\documentclass[sigconf, 10pt]{acmart}
\AtBeginDocument{%
  }

\setcopyright{none}
\copyrightyear{2018}
\acmYear{2018}
\acmDOI{XXXXXXX.XXXXXXX}
\acmConference[MobiSys '26]{The 24th ACM International Conference on Mobile Systems, Applications, and Services}{June 21 -- 25, 2026}{Cambridge, UK}
\acmISBN{978-1-4503-XXXX-X/2018/06}

\usepackage{booktabs}   
\usepackage{multirow}   
\usepackage{siunitx}  
\usepackage{placeins}
\usepackage{comment}
\usepackage{tabularx}  %
\begin{document}

\title{Silicone Ethernet (SEth): a Nervous System \\for Robotic Touch}


\author{Mengyao Liu}
\email{mengyao.liu@kuleuven.be}
\affiliation{%
  \institution{DistriNet, Computer Science, KU Leuven}
  \city{3001 Leuven}
  \country{Belgium}
}

\author{Dag Malstaf}
\email{dag.malstaf@kuleuven.be}
\affiliation{%
  \institution{DistriNet, Computer Science, KU Leuven}
  \city{3001 Leuven}
  \country{Belgium}
}

\author{Jonathan Oostvogels}
\email{jonathan.oostvogels@kuleuven.be}
\affiliation{%
  \institution{DistriNet, Computer Science, KU Leuven}
  \city{3001 Leuven}
  \country{Belgium}
}

\author{Sam Michiels}
\email{sam.michiels@kuleuven.be}
\affiliation{%
  \institution{DistriNet, Computer Science, KU Leuven}
  \city{3001 Leuven}
  \country{Belgium}
}

\author{Alexander Badri-Spröwitz}
\email{alexander.badri-sproewitz@kuleuven.be}
\affiliation{%
  \institution{RAM, Mechanical Engineering, KU Leuven}
  \city{Leuven}
  \country{Belgium}
}

\author{Danny Hughes}
\email{danny.hughes@kuleuven.be}
\affiliation{%
  \institution{DistriNet, Computer Science, KU Leuven}
  \city{3001 Leuven}
  \country{Belgium}
}

\renewcommand{\shortauthors}{Liu et al.}

\begin{abstract}
Fine-grained robotic touch sensing is essential for tasks such as robot-human interaction and the handling of hazardous materials. Yet, the sense of touch of robots is limited by the cost and complexity of routing cables to embedded sensors. This paper tackles this problem by contributing Silicone Ethernet (SEth), a wireless solution for touch sensing, communication and power transfer within a conductive silicone substrate. SEth~\emph{neurons} require no battery and deliver computation, communication, sensing and energy harvesting within a compact package. These neurons are placed into an undifferentiated conductive silicone substrate which may form the entire body of a soft robot, or an outer `skin' for hard robots. Our evaluation shows that SEth achieves data rates of 100\,kbps with sub-$\mu$W receive and mW-scale transmit power. Exploiting the unique properties of the conductive silicone substrate, SEth provides prioritized traffic arbitration similar to that found in wired control networks such as CAN. The SEth network inherently supports capacitive touch and presence sensing and neurons can harvest sufficient energy to transmit 10s of messages per second at a range of 1\,m. Considered in sum, these features open new degrees of freedom in touch sensing for soft robots.
\end{abstract}



\begin{teaserfigure}
  \centering
  \includegraphics[width=0.80\textwidth]{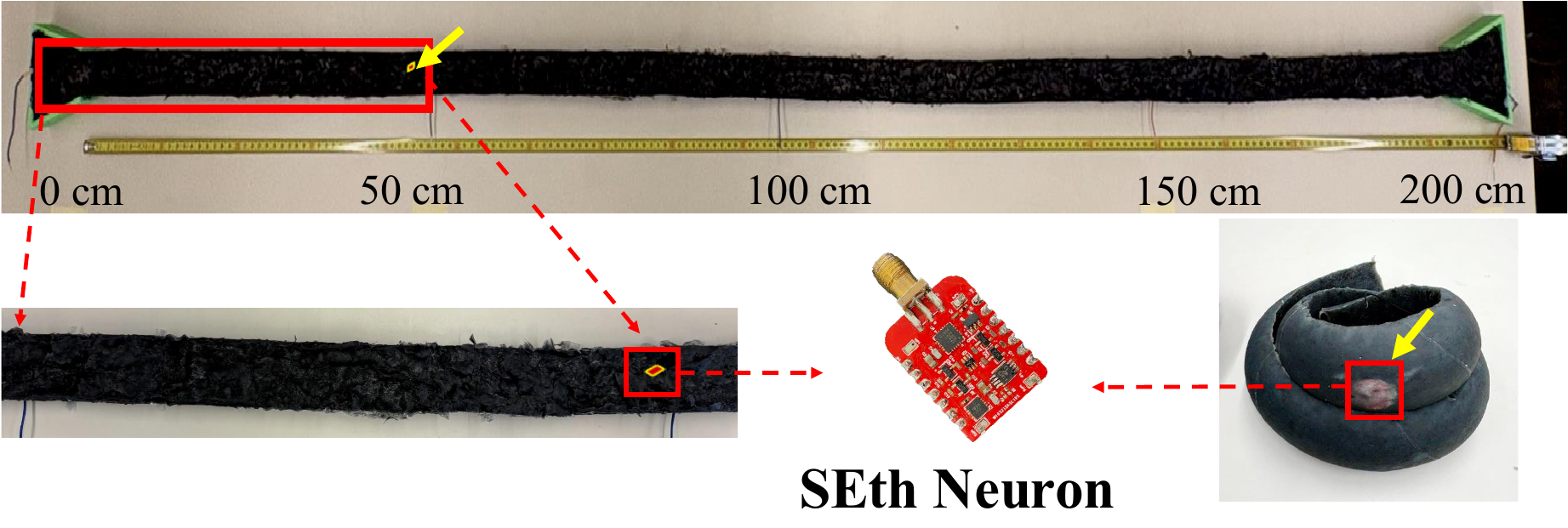}
  \caption{SEth unifies sensing, communication, and energy transfer for soft robots by introducing a capacitively coupled bus network embedded within a somewhat-conductive silicone substrate.}
  \label{fig:teaser}
\end{teaserfigure}


\maketitle
\section{Introduction}
\label{sec:introduction}
Touch is essential for modern robots, enabling them to explore and interact with their environments~\cite{8272325}. This is particularly important where robots are dealing with delicate materials as in fruit harvesting~\cite{9829727}, remote surgery~\cite{cianchetti2018biomedical} and human-robot interaction~~\cite{9561898}. While critical, delivering fine-grained touch sensing remains difficult due to the cost, complexity and mechanical limitations arising from routing power and network cables to large numbers of sensors that are distributed across the surface of the robot. State of the art approaches to tackling this problem focus on flexible Printed Circuit Boards (PCBs)~\cite{liu2022neuro} and lightweight wiring solutions~\cite{loher_stretchable_2006,rogers_materials_2010} that, at best, partially mitigate the problem. On the other hand, contemporary wireless solutions such as Bluetooth or IEEE 802.15.4~\cite{802154} fail to guarantee the deterministic Quality of Service (QoS) that real-time safety and critical systems such as robotics require~\cite{8057758, 802154}. In this paper, we advocate for a radical new approach that is based on wireless communication and power transfer over a conductive silicone substrate.

SEth (Silicone Ethernet) is built upon the concept of `neurons', compact and battery-free hardware modules that combine sensing, wireless communication and computation. These neurons are dropped into a conductive silicone substrate (the robot's `skin'), wherein they form a capacitively coupled network that provides efficient wireless power transfer at ranges of up to 2 meters, application data rates of up to 100 kbps and capacitive touch and presence sensing. SEth is a~\emph{real-time network} in the sense that it provides deterministic latency and prioritized traffic arbitration. The SEth transceiver can actively listen at orders of magnitude lower power than a traditional wireless transceiver, which eliminates the need for radio duty cycling and drives down latency. 


Considered as a whole, the SEth approach, for the first time, enables dense arrays of wireless touch sensors to be embedded within soft robotic skins, supports long-term operation without recharging or replacement and greatly simplifies mechanical design. The scientific contributions of this paper are three-fold:
\begin{enumerate}
    \item A novel approach to robotic presence and touch sensing based upon collaborating networks of wireless sensors. 
    \item An ultra-low power wireless transceiver with data rates of 100 kbps and realtime guarantees via prioritised arbitration.
    \item Efficient wireless power transfer techniques for conductive silicone substrates.
\end{enumerate}

The remainder of this paper is structured as follows. Section~\ref{sec:background} provides background on robotic touch sensing. Section~\ref{sec:design} describes the design of SEth, a novel sensing system for robots. 
Section~\ref{sec:implementation} describes the implementation of the SEth prototype.
Section~\ref{sec:evaluation} evaluates the performance of SEth under realistic conditions. 
Section~\ref{sec:relatedwork} discusses related work. Section~\ref{sec:conclusion} concludes. Finally, Section~\ref{sec:futurework} discusses directions for future work. 

\section{Background}
\label{sec:background}
This section provides background on robotic touch sensing, communication and power distribution in Sections~\ref{subsec:bg-touch} to~\ref{subsec:bg-power}. Based upon this, we distill requirements for SEth in Section~\ref{subsec:requirements}.

\subsection{Robot Touch Sensing}
\label{subsec:bg-touch}
Robotic touch is implemented using arrays of embedded sensors building upon capacitive, resistive, piezoelectric, or optical sensing principles~\cite{deanleon_wholebody_2019,ruppert_foottile_2020,sun_soft_2022}.
The use of touch sensors has been steadily expanding, and now includes applications not only in general robotics, but also in related domains such as prosthetics, rehabilitation, and haptics~\cite{kuchenbecker_characterizing_2003,dietrich_development_2004,choi_vibrotactile_2013,lee_nanomesh_2020}.
Typically, touch sensors are embedded on the surface of end effectors such as grippers, enabling robots to grasp fragile objects, detect contact and slip, manipulate tools, and interact with each other~\cite{mittendorfer_humanoid_2011}. However, conventional \emph{tactile} systems require wiring, onboard power, or bulky electronics, which complicates their integration, especially in soft robotic platforms that rely on compliant, stretchable materials like silicone rubber \cite{rus_design_2015}. Touch-sensors that combine soft membrane deformation with cameras or pressure sensors to reconstruct interaction forces \cite{ruppert_foottile_2020,sun_soft_2022} do not interfere with the soft membrane itself, but introduce strong constraints on sensor design, shape, and power- and communication routing, and therefore limit the multi-functionality of the soft structure. 

Wireless touch sensing systems are therefore desirable as they eliminate the need for communication cables, simplifying manufacture and dramatically improving mechanical flexibility, rigidity and weight. Eventually, based on these improvements, we expect a significant increase of the overall design space of touch sensing systems, and the emergence of new technologies. However, truly wireless touch sensing is difficult to achieve due to the need for real-time communication and the complexities of power distribution.

\subsection{Communication}
\label{subsec:bg-communication}
Robots often depend upon deterministic low latency control loops, wherein delays or jitter can cause degraded performance or even unsafe operation. This motivates the use of wired industrial protocols such as EtherCAT~\cite{EtherCAT} and CAN~\cite{ISO11898} that are capable of guaranteeing predictable timing, millisecond scale latency, and reliable synchronization across many distributed nodes.

While the flexibility of wireless communication is appealing, contemporary technologies such as Bluetooth~\cite{mockel_easy_2007}, Wi-Fi~\cite{IEEE802.11}, and 802.15.4~\cite{IEEE802.15.4} are not optimized for hard real-time determinism. Their Medium Access Control (MAC) protocols contend for the channel, backoff and retry cycles increase latency and jitter. Packet losses, channel congestion, and coexistence with other radio systems further compromise timing reliability. As a result, while traditional wireless options simplify cabling, they cannot guarantee the consistent, millisecond-scale timing that robotic sensing demands; motivating the need for~\emph{real-time} wireless network protocols.

\subsection{Power Distribution}
\label{subsec:bg-power}
Robotic systems typically distribute power to sensors and actuators through dedicated cables~\cite{grimminger_open_2020} or discrete batteries embedded within sensing modules~\cite{knight_energy_2008}. Dedicated power cables simplify energy management. However, physical wiring increases system complexity, adds weight and introduces potential points of mechanical failure due to bending fatigue and motion range limits. In the case of conventional robots, it requires dedicated cable routing solutions~\cite{sprowitz_roombots_2014,grimminger_open_2020}, especially in soft robots \cite{lu_flexible_2014}.

Battery-based sensors and actuators offer greater placement flexibility and eliminate the need for power cabling, enabling modular robotic architectures~\cite{sprowitz_roombots_2014} and easier integration into deformable or hard-to-reach surfaces~\cite{wang_biomorphic_2020}. This approach is well suited for distributed sensing systems that require independence from a central power bus. However, the drawbacks are significant; batteries increase weight, size, and require regular charging  or replacement. These disadvantages have so far prevented the widespread adoption of battery powered sensors and actuators; motivating research into wireless power transfer and battery-free designs.

\subsection{Requirements}
\label{subsec:requirements}
We argue for a radical new approach to robotic touch sensing that combines wireless power transfer and real-time networking with integrated sensing and communication to deliver a truly wireless and maintenance-free solution. SEth nodes can be easily embedded into robots, prosthetics, and haptic devices. Taking into account the background discussed above, we highlight the following four requirements:
\begin{enumerate}
    \item \emph{Wireless power transfer} provides a means to eliminate both power cables and high-maintenance batteries, facilitating the dense deployment of networked touch sensors.
    \item \emph{Real-time wireless networking} techniques are required, to support high-speed and deterministic sensing and actuation, as is expected from conventional wired solutions. 
    \item \emph{Integrated Sensing and Communication} provides a means to eliminate redundant components and thereby reduce the size of touch sensors.
    \item \emph{Low cost and complexity} are essential to maximize applicability and, in the longer term, a path towards a miniature Application Specific Integrated Circuit (ASIC).
\end{enumerate}

\section{Design}
\label{sec:design}

\begin{figure}
    \centering
    \includegraphics[width=1\linewidth]{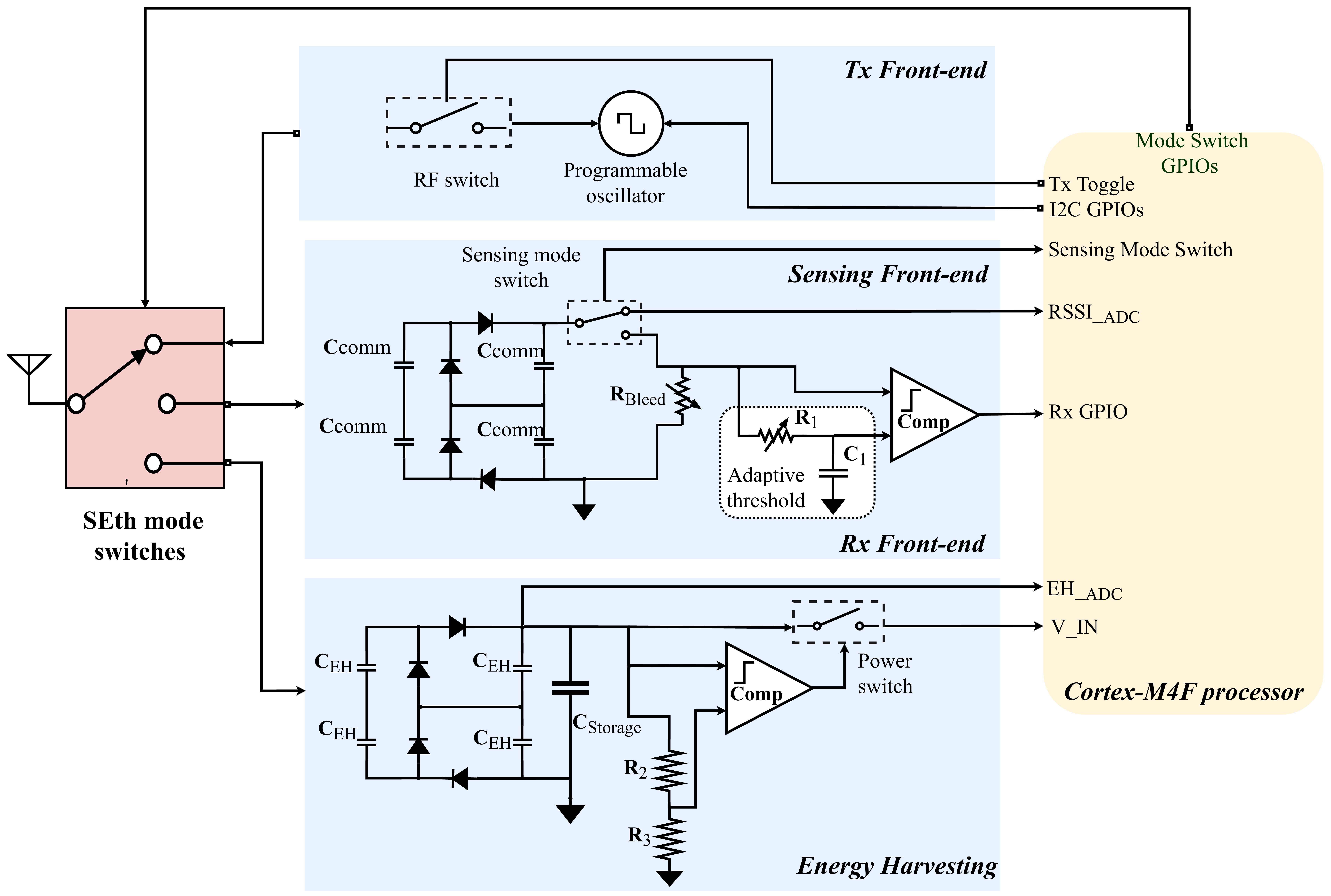}
    \caption{Hardware Block Diagram of the SEth Neuron}
    \label{fig:Hw_diagram}
\end{figure}
This Section describes the design of SEth, a system which aims to meet the requirements identified in Section~\ref{subsec:requirements}. We begin in Section~\ref{subsec:medium} by describing the conductive silicone substrate in which Neurons are deployed. Section~\ref{subsec:phy} describes the hardware design of the SEth neuron. Section~\ref{subsec:modes} describes the configuration of SEth in transmit, receive, harvesting and sensing modes. Finally Section~\ref{subsec:mac} describes the prioritized preemptive Medium Access Control (MAC) protocol of SEth.

\subsection{Network Medium}
\label{subsec:medium}

Our transmission medium is~\emph{polysiloxane}, commonly known as `silicone rubber', which we doped with ground-up carbon fiber for custom-level conductivity\footnote{Soft Robotics Toolkit contact sensor design: \url{https://softroboticstoolkit.com/contact-sensor/design}}. Specifically, we mixed 850\,g of Part~A and 850\,g of Part~B of a platinum-cure silicone  
(Ecoflex~00-20, Smooth-On\footnote{Ecoflex 00-20 product page: \url{https://www.smooth-on.com/products/ecoflex-00-20/}}) 
with 35\,g of 3\,mm ground carbon fiber that had been dispersed in 15\,g of 70\% isopropyl alcohol, which was drained off before doping. The mixture was left to cure in a 2\,m-long rectangular with a 2\,cm~\(\times\)~4\,cm cross-section mould at room temperature for 6\,h.  We measured an end-to-end resistance of approximately $1.7\,\mathrm{k}\Omega$ for the resulting 2\,m silicone bar, which is shown in Figure \ref{fig:teaser}. 

Soft parts made from this form of silicone rubber are common in robotics, haptics, rehabilitation, and medicine \cite{dietrich_development_2004,hepp_novel_2022}; various levels of shore hardness are available `off-the-shelf' and the material is highly durable, largely waterproof and can be doped and mixed with various materials to achieve custom properties. These types of parts are easy to manufacture by moulding, potentially supported by 3D printing. Silicone has become the quasi-standard material of soft robots~\cite{rus_design_2015}, and is a common choice for medical and haptics devices due to being bio-compatible and bio-inert \cite{dietrich_development_2004}.

\begin{figure}
    \centering
    \includegraphics[width=1.0\linewidth]{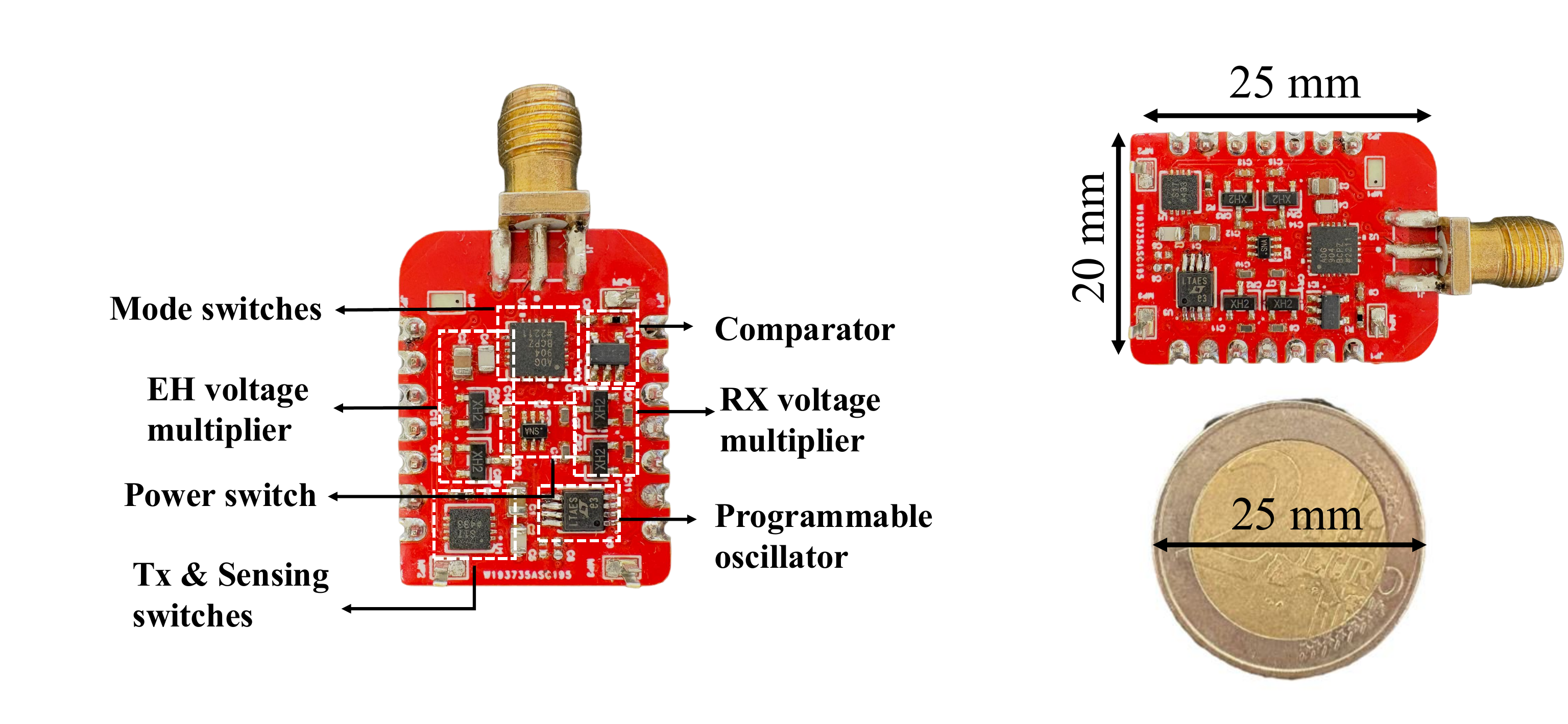}
    \caption{The SEth neuron prototype, showing the key components on the 25\,mm~\(\times\)~20\,mm PCB. A 2\,€ coin is included for scale.}
    \label{fig:SEth prototype}
\end{figure}

\subsection{The Neuron PHY Layer}
\label{subsec:phy}
The hardware design of the SEth neuron is shown in Figure~\ref{fig:Hw_diagram}. It is composed of five communicating hardware modules; a switchable antenna, the transmitter front-end, receiver front-end, energy harvester and network processor. These elements are described in Section~\ref{subsubsec:design-antenna} to~\ref{subsubsec:design-mcu}. Figure \ref {fig:SEth prototype} shows an annotated view of the SEth neuron prototype.


\subsubsection{Switchable Antenna}
\label{subsubsec:design-antenna}
SEth does not require  an optimized antenna like a standard far-field radio, which would anyway be infeasible in the VHF range with wavelengths from 1 to 10\,m. All that is required is a small electrode which makes contact with the conductive silicone medium. This antenna is switchable, enabling the neuron to move between energy harvesting, transmit, receive and sensing functionality by connecting it to the appropriate hardware module, as described in Section~\ref{subsec:modes}. By default, the switch connects the antenna to the neuron's energy harvesting module, enabling charging in cases where the energy storage capacitor is depleted and the application processor is powered down. It should be noted that the SMA connector is not necessary for this medium and is included for other applications of the transceiver.

\subsubsection{Transmitter (TX) Front-End}
\label{subsubsec:design-tx}

The SEth transmitter design is kept simple, using an ON/OFF Keyed (OOK) baseband signal provided by the network processor. 
The VHF carrier wave is generated by an I2C-controlled programmable oscillator (Analog Devices LTC6904) with a frequency range from 1\,kHz to 68\,MHz%
\footnote{LTC6904 datasheet: \url{https://www.analog.com/en/products/ltc6904.html}}.
This provides flexibility and has allowed us to apply the same transceiver in free space  ~\cite{liu2024cain}, on body ~\cite{liu2025enoch}, and now in conductive silicone. 
The carrier frequency was selected as 64\,MHz based on empirical performance experiments across the available frequency range. 
The transmitter has a maximum power consumption of 15.9\,mW at 64\,MHz. 
While this simple modulation approach is sub-optimal in terms of spectrum usage, it enables the use of an extremely simple and low-power receiver, as explored in the following section.

\subsubsection{Receiver (RX) Front-End}
\label{subsubsec:design-rx}
The SEth receiver front-end is an extremely low-power design inspired by wake-up radios that extends the ENOCH design that we presented in~\cite{liu2025enoch}. The front end features passive amplification and rectification of the incoming RF signal using a two-stage Greinacher voltage multiplier.  This transforms the signal into a form that can be processed in the digital domain without the need for high power Analogue-to-Digital Converters (ADCs). The OOK signal is decoded by a digital comparator in combination with an RC low pass filter which performs dynamic thresholding of the input signal and increase immunity against background noise. Receiver gain is controlled through a variable bleed resistor \textit{R2}, which prevents receivers being swamped by powerful transmitters. A complete list of components and their sizing is provided in Section~\ref{sec:implementation}.

\subsubsection{Energy Harvester}
\label{subsubsec:design-eh}
The Energy Harvesting sub-system shares the same front-end architecture as the receiver, except that rectified RF energy is used to charge a large storage capacitor in the range of 100\,$\mu$F to 500\,$\mu$F. Despite this similarity, the optimal sizing of capacitors for power transfer and communication differ by orders of magnitude~\cite{static}, making it difficult to serve both purposes with a common circuit. Furthermore, separate front ends allow us to use purpose-built RF capacitors on the communication module. The energy harvester also embeds a power isolation switch which uses a low-power comparator and digital switch to keep the network processor off until sufficient power has been harvested to support communication. Technically, the comparator is out-of-spec. when the capacitor voltage falls below 1.8V, however, the switch is default off and its power consumption remains in the sub-$\mu$W range (compared to mW for the network processor in brown-out). When active, the network processor can monitor the state of charge directly via its ADC to inform scheduling decisions.

\subsubsection{Network Processor}
\label{subsubsec:design-mcu}
The SEth software stack is hosted on an Ambiq Apollo3 Blue MCU\footnote{\href{https://ambiq.com/apollo3-blue/}{Ambiq Apollo 3 Blue MCU, product information available at https://ambiq.com/apollo3-blue/}}, which offers a single-core ARM Cortex-M4F core clocked at 48\,MHz, 384\,kB of RAM, 1\,MB of flash, and a 14-bit ADC. External peripherals may be connected via SPI, I\textsuperscript{2}C, or UART. Critically for our application, the processor is extremely low power,  with sub-$\mu$W sleep modes and an active power consumption of approximately $6\,\mu\text{A}/\text{MHz}$. Once booted by the power isolator subsystem, the processor configures the neuron to operate according to the modes outlined in Section~\ref{subsec:modes}. Developing an end-to-end protocol capable of orchestrating large networks of neurons is the subject of our future work.

\subsection{Modes of Operation}
\label{subsec:modes}
This section summarizes the transmitter and receiver configurations for energy harvesting, communication and sensing.
\begin{enumerate}
    \item \textbf{Energy Harvesting:} The transmitter enables its TX front-end, generating a 64\,MHz sine wave while the network processor idles. The receiver enables its harvesting front end to accumulate charge in its storage capacitor, while the network processor is powered-down or in sleep mode. We expect that power transmission will be performed by an externally powered \emph{coordinator} node, though its functionality is otherwise undifferentiated from a standard neuron.
    \item \textbf{Communication:} The transmitter enables its TX front-end, generating a 64\,MHz carrier wave, which the network processor mixes with the coded base-band through On/Off Keying. The receiver enables its RX front-end and tailors the bleed resistor configuration to match incoming signal strength as measured by its ADC. The received signal is demodulated and passed to the network processor via GPIO. 
    \item \textbf{Sensing:} The transmitter enables its TX front-end, generating a 64\,MHz sine wave while the network processor idles. The receiver enables its RX front-end and sensing mode switch, so that the sensing signal is sampled by the MCU's ADC and processed in software.
\end{enumerate}


\subsection{Medium Access Control}
\label{subsec:mac}
SEth Neurons control access to the wireless medium using the prioritized and preemptive arbitration scheme that we proposed in EnOCH~\cite{liu2025enoch} and that is itself inspired by CAN bus. To avoid repetition, the complete algorithm is not reproduced here, however, from a high level it operates as follows:

\begin{enumerate}
    \item The node checks whether data frames are available for transmission. If no data is available, SEth continues to listen for incoming data. 
    \item When data frames are available and the channel turns idle, SEth switches to TX mode and transmits a priority preamble, signaling the system's intent to take control of the conductive silicone medium. 
    \item After the priority preamble, the system activates the receiver front-end and reevaluates the channel's status. If the channel is occupied, the system infers there is a competing transmission with a higher-priority preamble, causing it to revert to receive mode, awaiting the next available transmission opportunity. If the channel remains available, the system proceeds to transmit the remainder of the data frame.
    \item Once the frame's transmission is complete, SEth reverts to receive mode. This concludes a full  transmission cycle.
\end{enumerate}

\paragraph{Frame structure}
The data link layer frame format of SEth consists of the following parts.
\begin{itemize}
    \item A priority preamble, which consists of 1 to  $N_{priority}$ repetitions of a 10\,µs \texttt{on} and a 10\,µs \texttt{off} symbol, where $N_{priority}$ is a configurable parameter that determines the number of distinct priority levels in the system; longer preambles correspond to higher priority levels. The priority level associated with a frame is decided in software at transmission time. 
    \item A 0.2\,ms start of frame delimiter, consisting of the same on-off sequence as the priority preamble.
    \item An 8-bit destination address.
    \item A 48-bit payload  containing free-form application data.
    \item An 8-bit Cyclic Redundancy Check (CRC) to detect corrupt packets.
   \end{itemize}

\paragraph{Line Coding} 
SEth applies differential Manchester encoding~\cite{horowitz1989art} to its on-off keyed transmissions, using signal level transitions to encode bits. The additional redundancy provided by such coding increases the system's noise immunity. The self-clocking nature of differential Manchester encoding enables the synchronization and decoding of data without separate clock signals.

\section{Implementation}
\label{sec:implementation}

The SEth prototype is illustrated in Fig.~\ref{fig:SEth prototype}. Its dimensions are 25\,mm $\times$ 20\,mm
.~All operational logic, signal processing and the data link layer are implemented in software on the MCU. All firmware is written as bare-metal C code using the Visual Studio Code IDE. All software and hardware will be published under an open-source license upon acceptance of the paper. 


Table~\ref{tab:bomSEth} outlines the bill of materials for all components in a batch of 10,000 units. Even built from discrete components, SEth neurons have a relatively low cost and complexity. Furthermore, reflecting upon the bill of materials, it can be seen that SEth requires no crystal oscillator, battery or other exotic components, suggesting that there may be a route to a dramatically lower cost single-chip design.


\begin{table}[h]
    \centering
    \renewcommand{\arraystretch}{1.3} 
    \caption{Bill of materials for our SEth prototype }
    \label{tab:bomSEth}
    \begin{tabularx}{\linewidth}{
        >{\centering\arraybackslash}X
        >{\centering\arraybackslash}X
        >{\centering\arraybackslash}X}
        \hline
        \textbf{Component} & \textbf{Part number} & \textbf{Price (USD)} \\
        \hline
        Ambiq Apollo3 Blue MCU   & AMA3B2KK-KBR  & \$3.94 \\
        Rectifying diodes        & SMS7630-005LF & \$0.66 \\
        RF SP4T switch           & ADG904        & \$3.20 \\
        RF SPST switch           & ADG902        & \$2.37 \\
        Power switch             & SIP32431DR3   & \$0.23 \\
        Analog comparator        & TLV7031       & \$0.09 \\
        Programmable oscillator  & LTC6904       & \$3.80 \\
        RF capacitors            & 4.7 pF        & \$0.47 \\
        Resistors and capacitors & N/A           & \$0.27 \\
        \hline
        \textbf{Total}           &               & \textbf{\$15.03} \\ 
        \hline
    \end{tabularx}
\end{table}

\section{Evaluation}
\label{sec:evaluation}
Sections~\ref{subsec:eval-network} through~\ref{subsec:eval-touch} evaluate SEth with respect to network performance, contention resolution, wireless power transfer and touch sensing respectively. In all cases, the SEth transceiver was configured to operate at a carrier frequency of 64\,MHz, which was empirically determined to be optimal for the conductive silicone medium. Unless otherwise noted, experiments were conducted on our 2\,m conductive silicone testbed as shown in Figure~\ref{fig:teaser}.

\subsection{Network Performance}
\label{subsec:eval-network}
\textbf{Speed: }The SEth transceiver achieves a symbol rate of 200 kilobaud, which gives rise to an application data rate of 100\,kbps after Manchester coding is applied. The maximum data rate of SEth is limited by the speed at which the comparator can demodulate the OOK signal and was determined through incremental experimentation. 

\textbf{Latency: } Accounting for per-frame overhead, the SEth transceiver delivers 8-byte packets in 840\,$\mu$s. Crucially, the wake-up radio inspired design of SEth means that nodes are~\emph{always} responsive and listening for incoming packets, unlike conventional radios which must be duty-cycled due to their high listening power consumption, leading to either dramatically higher latency or power consumption~\cite{10.1145/1031495.1031508}. Control-loop latency is hence dramatically lower than that of popular radio systems such as IEEE~802.15.4~\cite{802154}.

\begin{figure}[t]
    \centering    \includegraphics[width=0.85\linewidth]{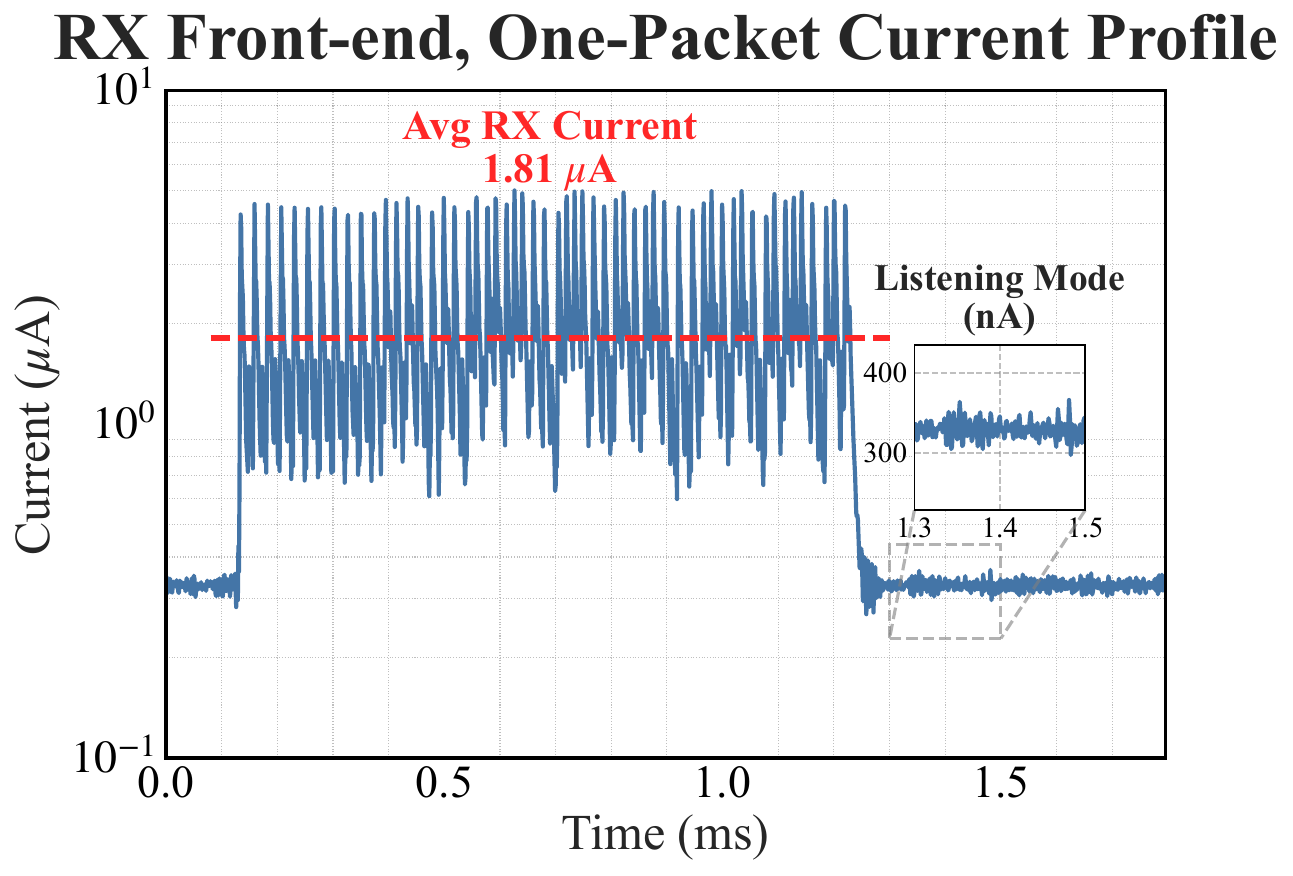}
    \caption{Power profile for RX front-end.}    \label{fig:rx_power_frontend}
\end{figure}

\begin{figure}[t]
    \centering
    \includegraphics[width=0.85\linewidth]{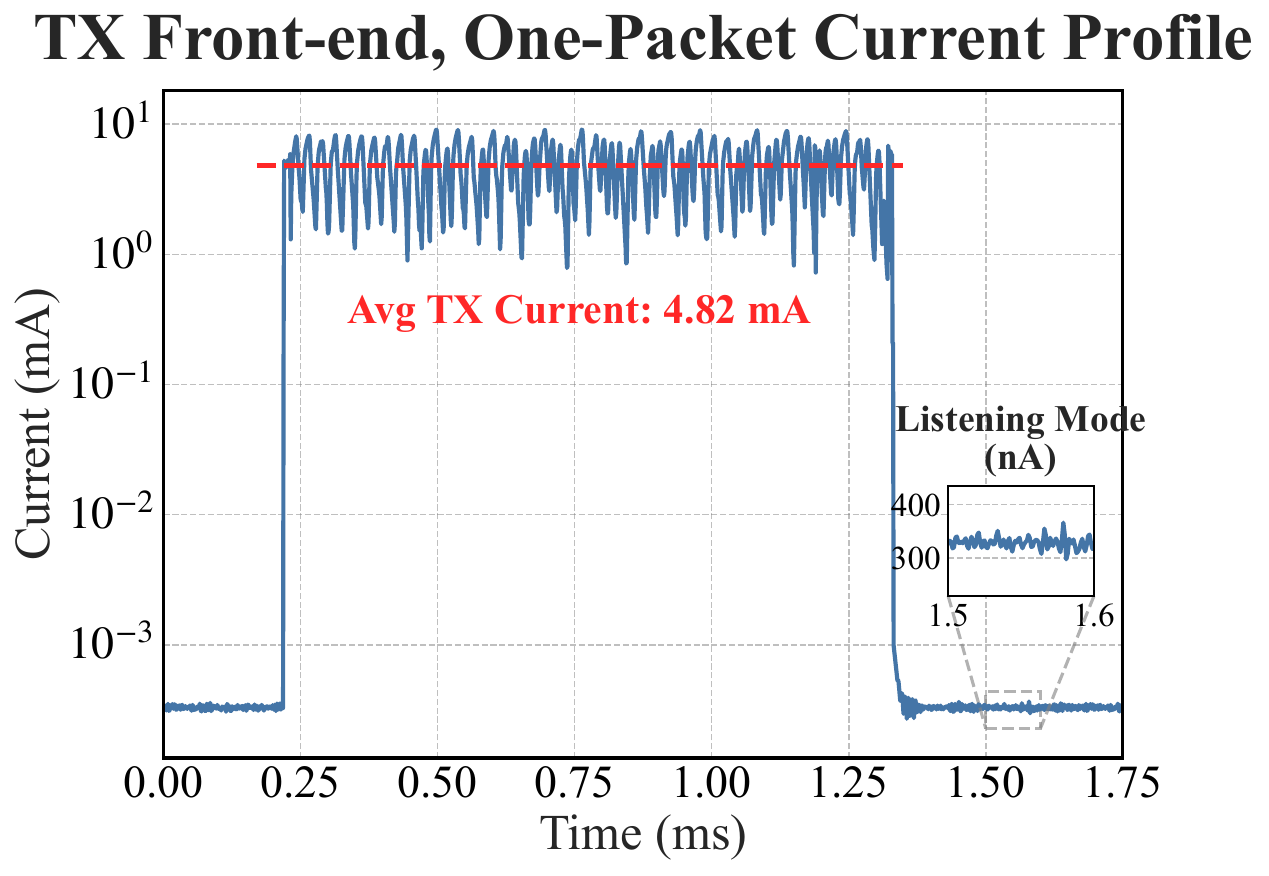}
    \caption{Power profile for TX front-end.}
    \label{fig:tx_power_frontend}
\end{figure}

\begin{figure}[t]
    \centering
    \includegraphics[width=0.85\linewidth]{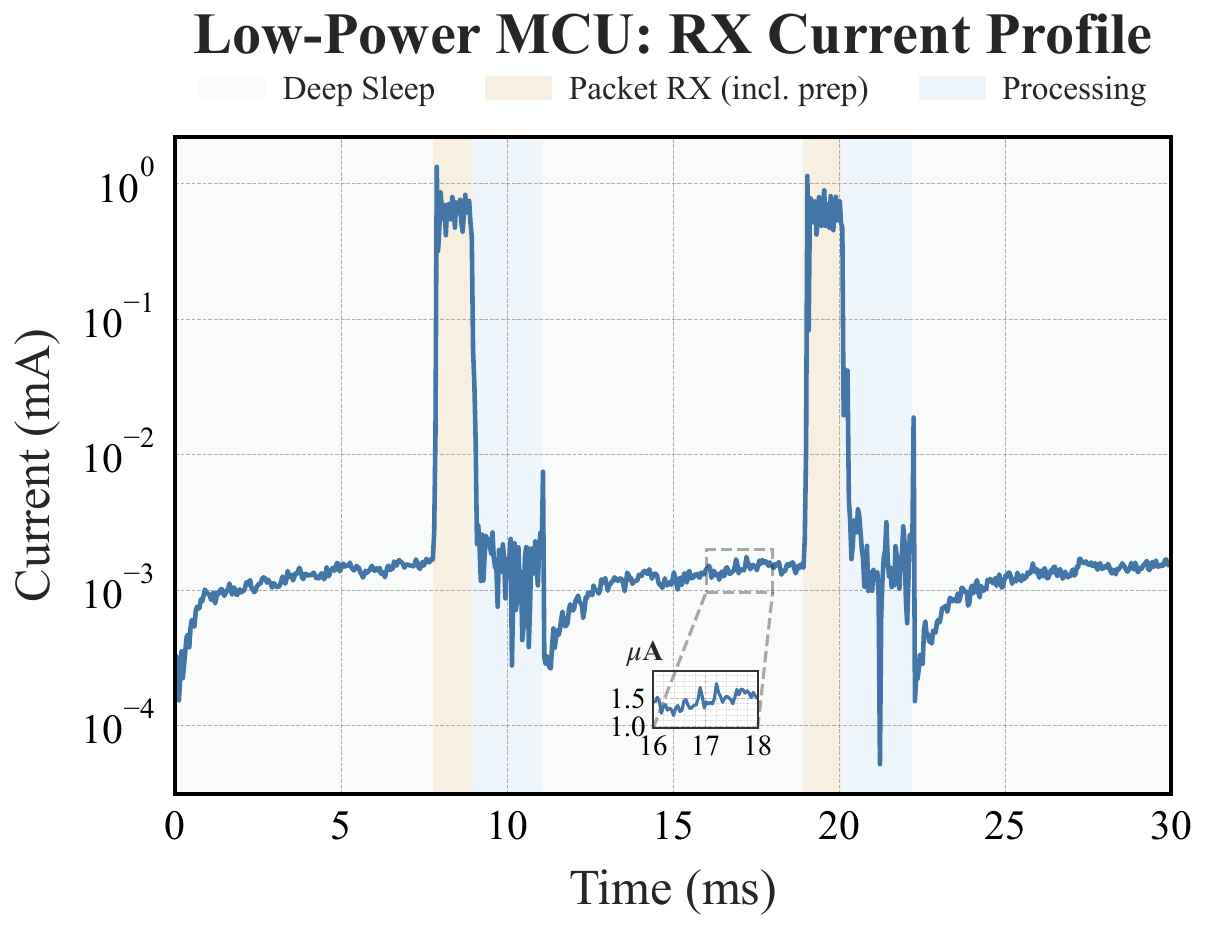}
    \caption{Power profile for MCU during RX.}
    \label{fig:rx_power_integrated}
\end{figure}

\begin{figure}[t]
    \centering
    \includegraphics[width=0.85\linewidth]{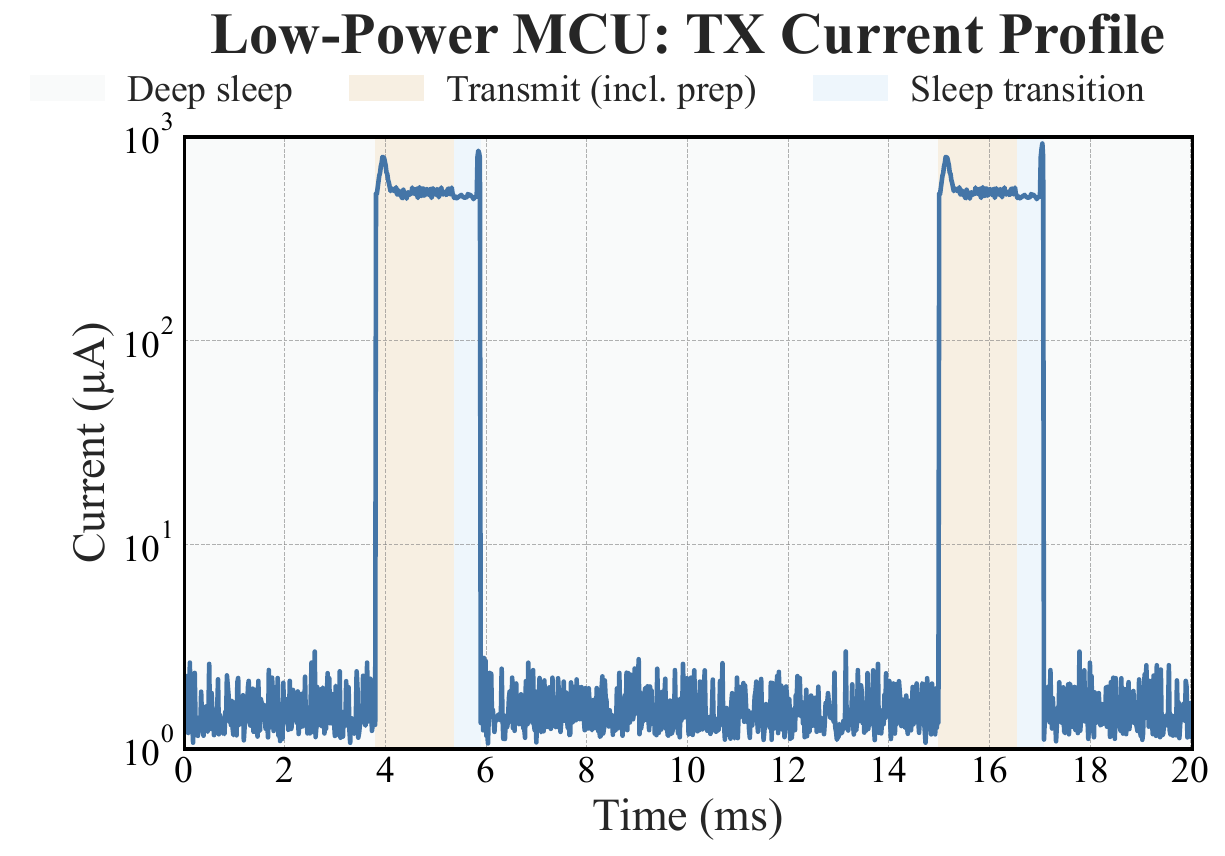}
    \caption{Power profile for MCU during TX.}
    \label{fig:tx_power_integrated}
\end{figure}

\textbf{Power: }Figure~\ref{fig:rx_power_frontend} and~\ref{fig:tx_power_frontend} isolate the power consumption of the SEth transceiver front-end while receiving and transmitting respectively. As can be seen from the figures, SEth is ultra low power, consuming an average of 320\,nA in listening mode with an average receiving current draw of 1.8\,$\mu$A and an average transmitting current draw of 4.82\,mA. Figure~\ref{fig:rx_power_integrated} and~\ref{fig:tx_power_integrated} show the same analysis for SEth's microcontroller, which mediates communication for repeated receive and transmit cycles. As can be seen from the figures, the microcontroller consumes an average of~1.5\,$\mu$A while listening, with an average current draw of 500\,$\mu$A when receiving and transmitting. This amounts to orders of magnitude lower power consumption than conventional radios. For example, the  Nordic nRF52840\footnote{\href{https://docs.nordicsemi.com/bundle/ps\_nrf52840/page/keyfeatures\_html5.html}{https://docs.nordicsemi.com/bundle/ps\_nrf52840/page/keyfeatures\_html5.html}} consumes on the order of 5\,mA while listening and receiving. We note that apparent transmission durations in the power graphs exceed the actual duration of a packet, as both MCU and radio front-end take time to switch between operational modes.

\begin{figure}[t]
    \centering    \includegraphics[width=0.85\linewidth]{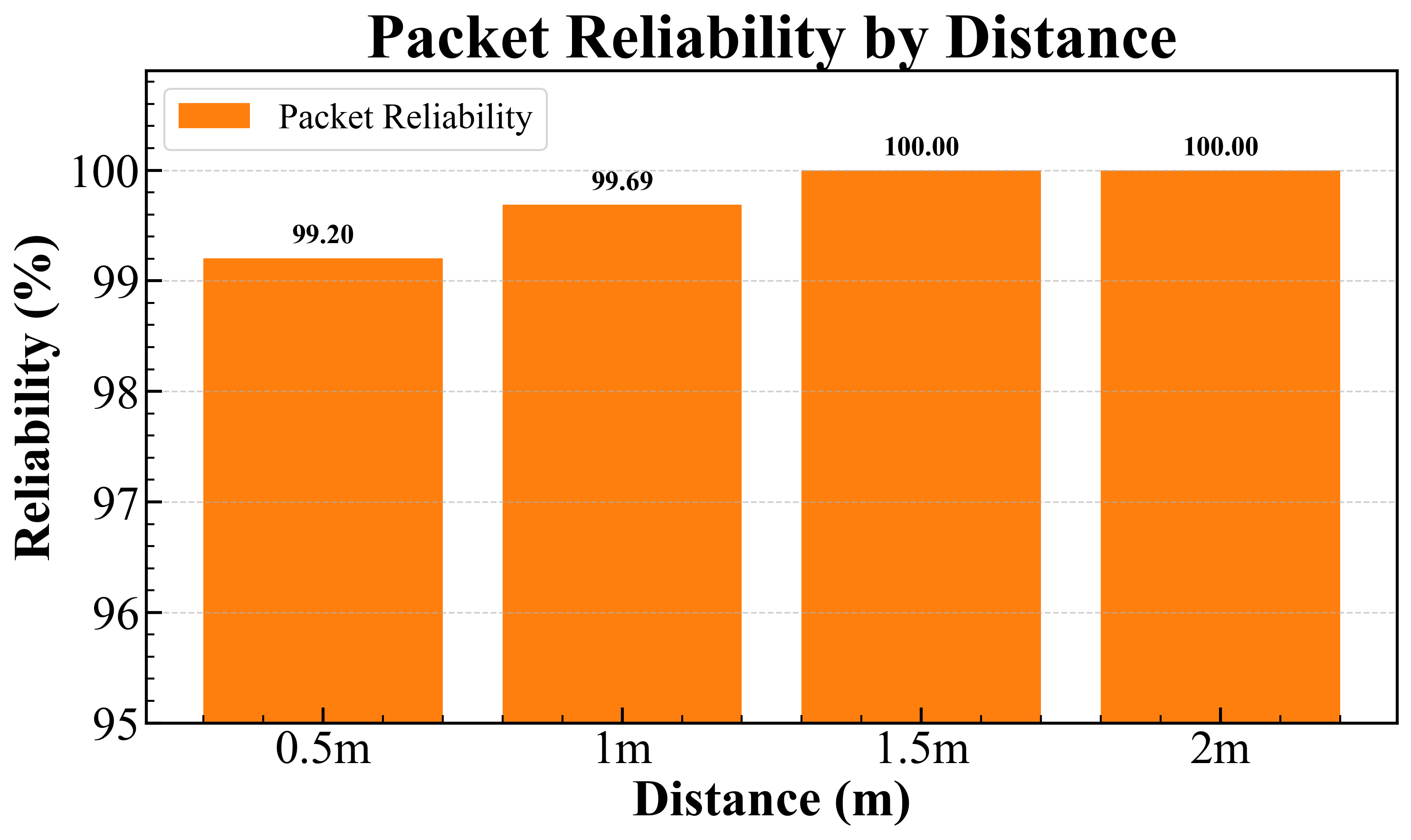}
    \caption{Communication reliability of SEth as a function of distance}
    \label{fig:reliability}
\end{figure}

\textbf{Reliability: }Figure~\ref{fig:reliability} shows the reliability of communication between a pair of SEth nodes for ranges between 50 and 200\,cm. As can be seen from the figure, SEth  delivers $>99$\% of packets correctly. There is no significant relationship between communication reliability and range within the silicone substrate. Transmissions that are not received correctly tend to have several tens of bit errors: higher-layer reliability techniques such as retransmissions will hence be required.


\subsection{Contention Resolution}
\label{subsec:eval-contention}
As described in Section~\ref{sec:design}, SEth supports real-time operation through non-destructive, prioritized contention resolution following the same model as CAN bus~\cite{farsi1999overview}. In this model, each node is assigned a unique priority code and, when two nodes attempt to access the channel at the same time, contention is resolved in favour of the highest-priority node, while the lower-priority transmission is suppressed. This approach minimises packet loss and ensures that higher priority traffic is delivered with lower latency. 

\begin{figure}[t]
    \centering    \includegraphics[width=1\linewidth]{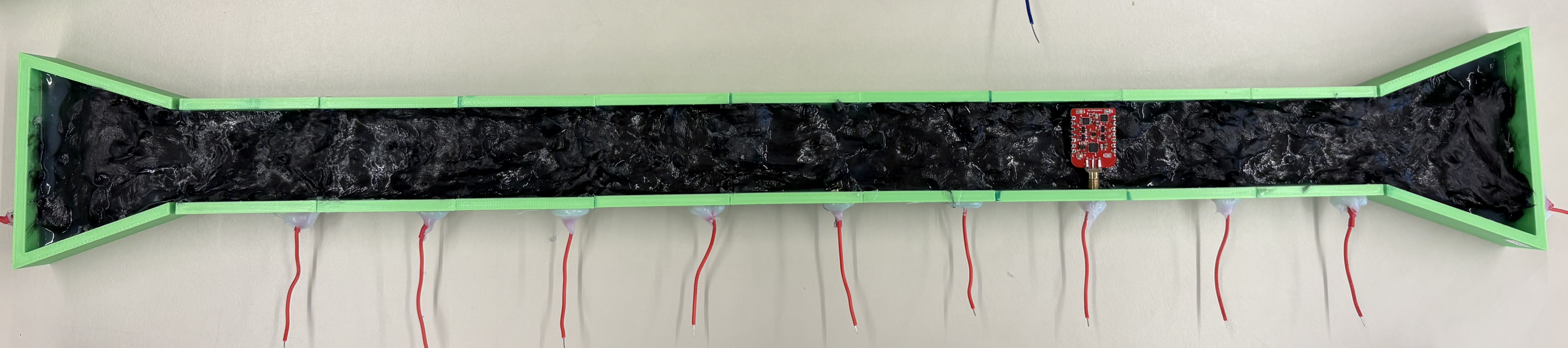}
    \caption{Experimental set-up for the evaluation of SEth's contention resolution mechanism. The set-up consists of 50 \,cm of conductive silicone. Each wire indicates the position of one SEth node.}
    \label{fig:contentionsetup}
\end{figure}

\begin{figure}[t]
    \centering
\includegraphics[width=0.95\linewidth]{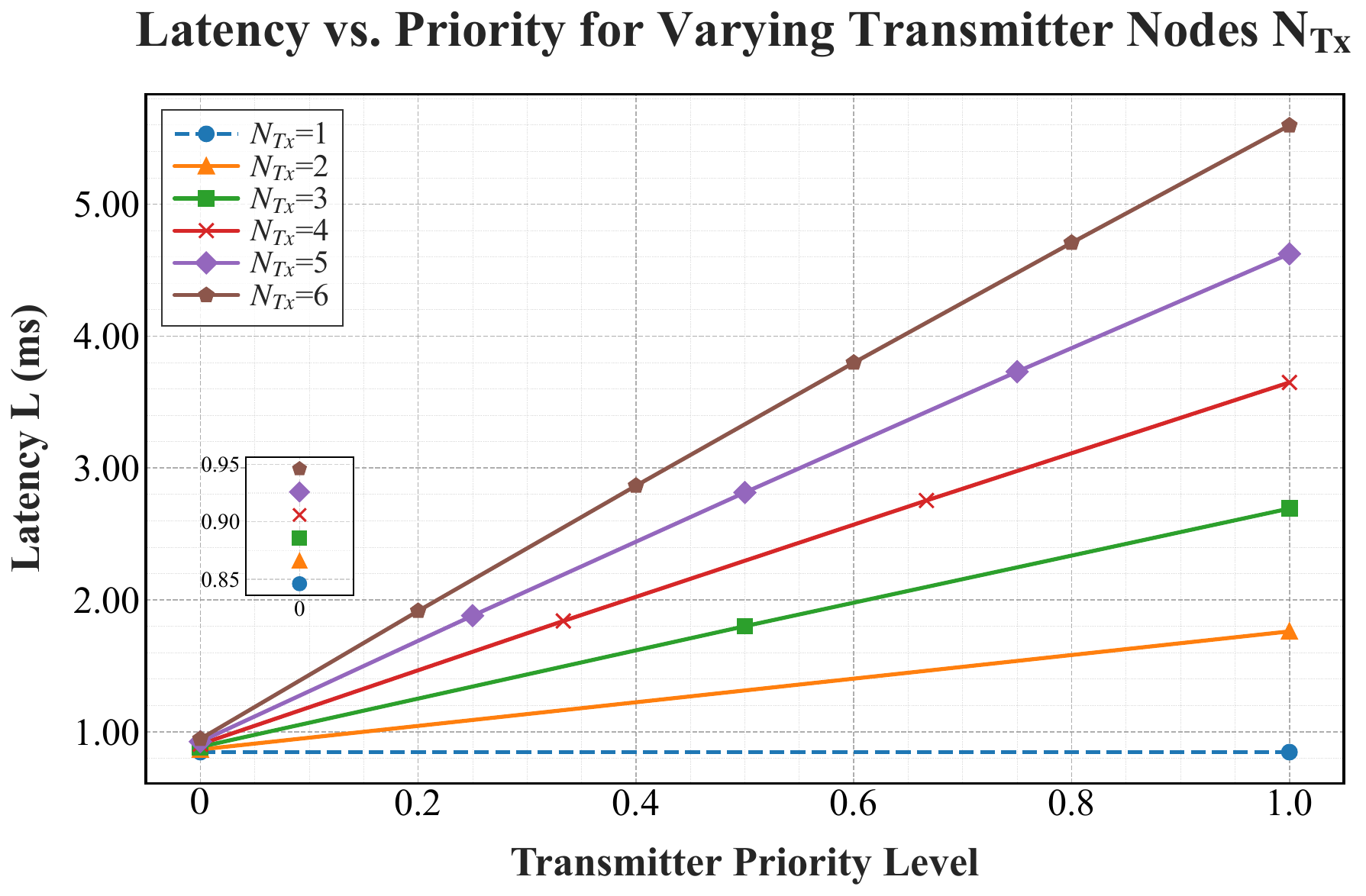}
    \caption{Latency of frame delivery under contention as a function of assigned priority level. $N_{tx}$ denotes the number of contenders in a given setting.}
    \label{fig:contention}
\end{figure}

Figure~\ref{fig:contention} evaluates  latency vs transmitter priority level for 6 nodes with unique priorities  represented as (discrete) values between 0 and 1. As can be seen from the figure, latency scales linearly with assigned priority. To assess contention resolution in the dense deployments, we evaluate it in a 50 cm section of conductive silicone with nine evenly spaced nodes, as shown in Figure~\ref{fig:contentionsetup}. For each experimental replicate, the leftmost node triggers contention by requesting an immediate reply from the first $N$ nodes to its right; these reply packets then contend for the channel.

\begin{figure}[t]
    \centering
\includegraphics[width=0.95\linewidth]{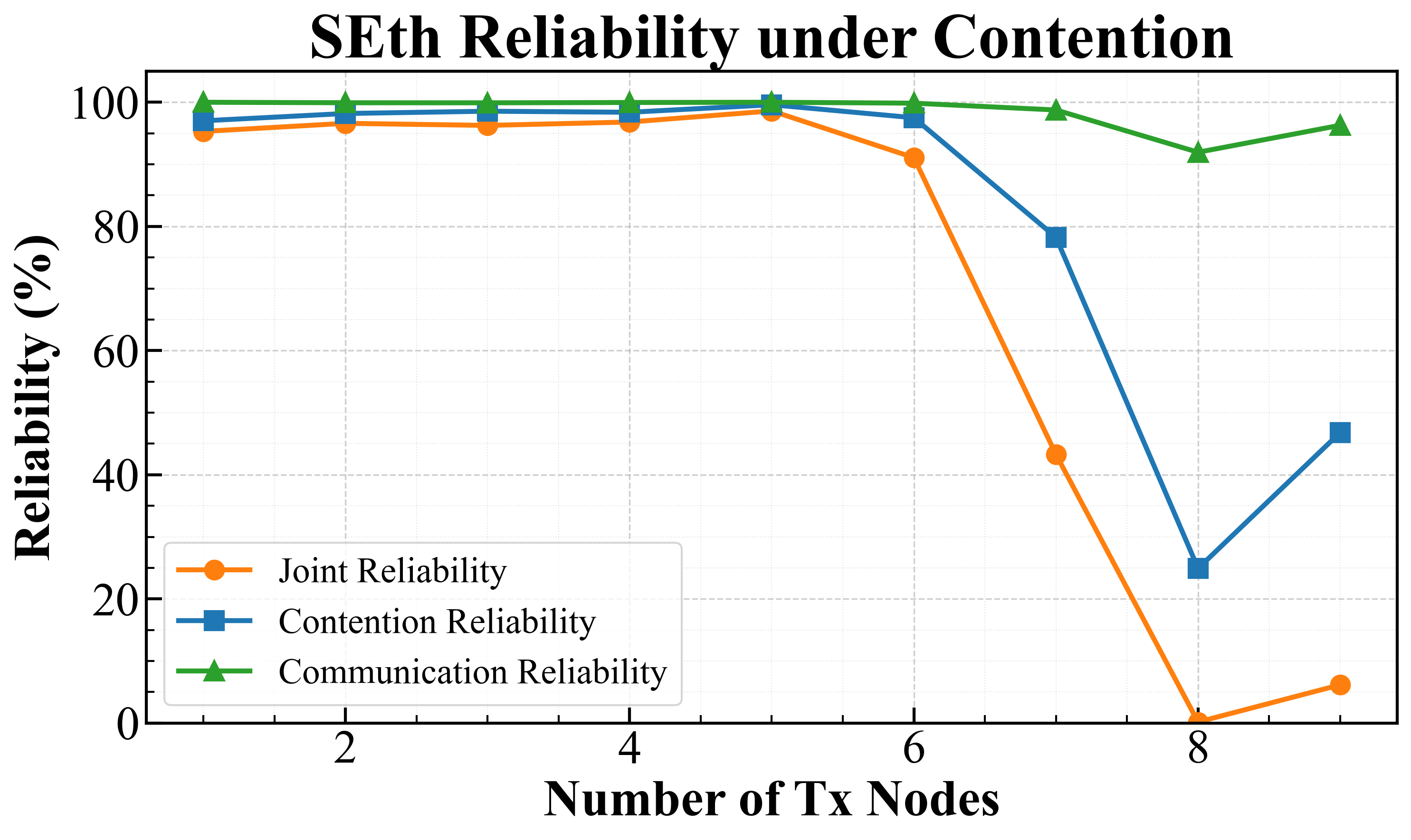}
    \caption{Reliability of SEth under contention.}
    \label{fig:pdr_plot}
\end{figure}

Figure~\ref{fig:pdr_plot} analyses the success rate of contention resolution and subsequent packet transmission. ``Contention Reliability'' indicates the likelihood that a received packet indeed originates from the highest-priority node, while ``Communication Reliability'' refers to the likelihood of receiving a correct frame, irrespective of the priority level of its transmitter. ``Joint Reliability' indicates the likelihood of a fully correct scenario, i.e.\ the conjunction of both events. As can be seen from the figure, contention has negligible impact on the probability of successful packet transmission from nodes which results in Communication Reliability exceeding 95\% in all scenarios. However the correctness of prioritized arbitration drops quickly for 6 or more contenders. We are currently investigating methods to raise the robustness of this mechanism. 


\subsection{Wireless Power Transfer}
\label{subsec:eval-wpt}
\begin{figure}[t]
    \centering
\includegraphics[width=1\linewidth]{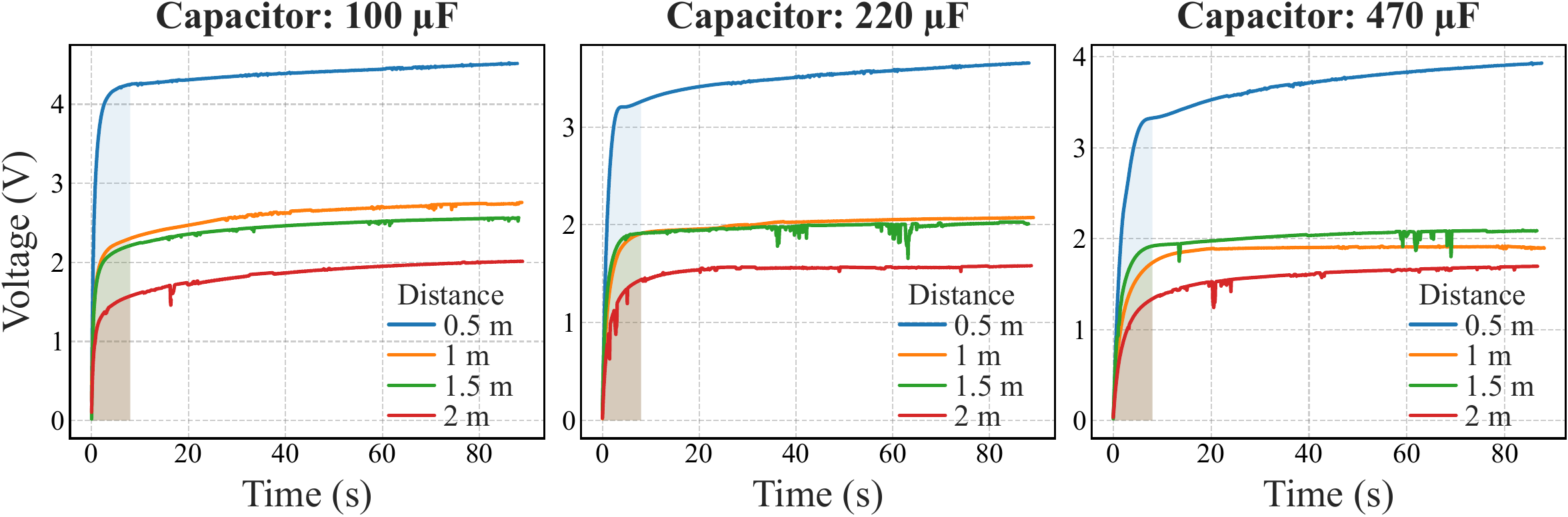}
    \caption{Wireless Power Transfer}
    \label{fig:power_transfer}
\end{figure}

As described in Section~\ref{sec:design}, a \emph{coordinator} neuron connected to external power is capable of providing sufficient Radio Frequency (RF) energy to power SEth Neurons. As described in Section~\ref{sec:design}, SEth neurons activate when the capacitor reaches 2\,V and brown-out at 1.8\,V. Figure~\ref{fig:power_transfer} shows the evolution of energy available to SEth neurons at ranges from 0.5\,m to 2\,m for capacitor sizes ranging from 100\,$\mu$F, 220\,$\mu$F and 470\,$\mu$F, starting from a state in which their energy storage capacitor is fully depleted. As can be seen from the figure, all capacitors charge to activation voltage at distances of 1.5\,m or less. However, at a distance of 2\,m, all but the 100\,$\mu$F capacitor fail to activate due to higher leakage current. We believe that this is not a major problem however, as the transmit power of the coordinator neuron can trivially be increased by an order of magnitude through the addition of a Low-Noise Amplifier (LNA). 

Once activated, neurons may continue to do work while harvesting energy. Tables~\ref{tab:duty-cycle} and ~\ref{tab:powerup} explore their performance in two dimensions by identifying the maximum sustainable packet rate that neurons can support without browning out, and quantifying the time necessary for initially charging a neuron to its 2\,V setpoint from a completely discharged state, respectively. This sustainable packet rate is determined by the time necessary for the transmission of a packet and the subsequent time necessary for the voltage on the energy storage capacitor to again reach its initial 2\,V operating point.  As can be seen from Figure~\ref{fig:power_transfer}, said capacitor can, especially at lower ranges, accumulate sufficient energy to enable the transmission of bursts that comprise tens of packets before brownout. \\


\begin{table}[t]
  \centering
  \caption{Number of packets that can be transmitted per second within the energy budget provided by SEth's energy transfer mode.}
  \label{tab:duty-cycle}
  \begin{tabular}{lccc}
    \toprule
    \multirow{2}{*}{Distance [m]} & \multicolumn{3}{c}{Capacitor size} \\
    \cmidrule(l){2-4}
     & 100\,\si{\micro\farad} & 220\,\si{\micro\farad} & 470\,\si{\micro\farad} \\
    \midrule
    0.5 & 27.3 & 17.7 & 24.6 \\
    1.0 & 10.6 & 5.6 & 10.1 \\
    \bottomrule
  \end{tabular}
\end{table}

\begin{table}[t]
  \centering
  \caption{Time in seconds required to charge a SEth node to its operating voltage from a fully discharged state.}
  \label{tab:powerup}
  \begin{tabular}{lccc}
    \toprule
    \multirow{2}{*}{Distance [m]} & \multicolumn{3}{c}{Capacitor size} \\
    \cmidrule(l){2-4}
     & 100\,\si{\micro\farad} & 220\,\si{\micro\farad} & 470\,\si{\micro\farad} \\
    \midrule
    0.5 & 0.42 & 1.36 & 1.82 \\
    1.0 & 1.02 & 2.79 & 3.43 \\
    \bottomrule
  \end{tabular}
\end{table}

\subsection{Touch Sensing}
\label{subsec:eval-touch}

Capacitive coupling not only transfers power and information, but has also been proposed as a sensing modality for soft robots. Below, we examine the extent to which such sensing can be unified with SEth's approach towards communication and energy transfer. We position two nodes at opposite ends of our 2-metre silicone testbed. One of these nodes continuously transmits a 64\,MHz carrier. The RSSI voltage of receiver's RX front-end
is continuously monitored using SEth's sensing mode, while an individual slowly steps towards the testbed with outstretched hand, moving perpendicularly to the testbed's central axis, eventually stopping when firmly squeezing the silicone material at some distance $L$ from the transmitting node.

\begin{figure}
    \centering
    \includegraphics[width=0.95\linewidth]{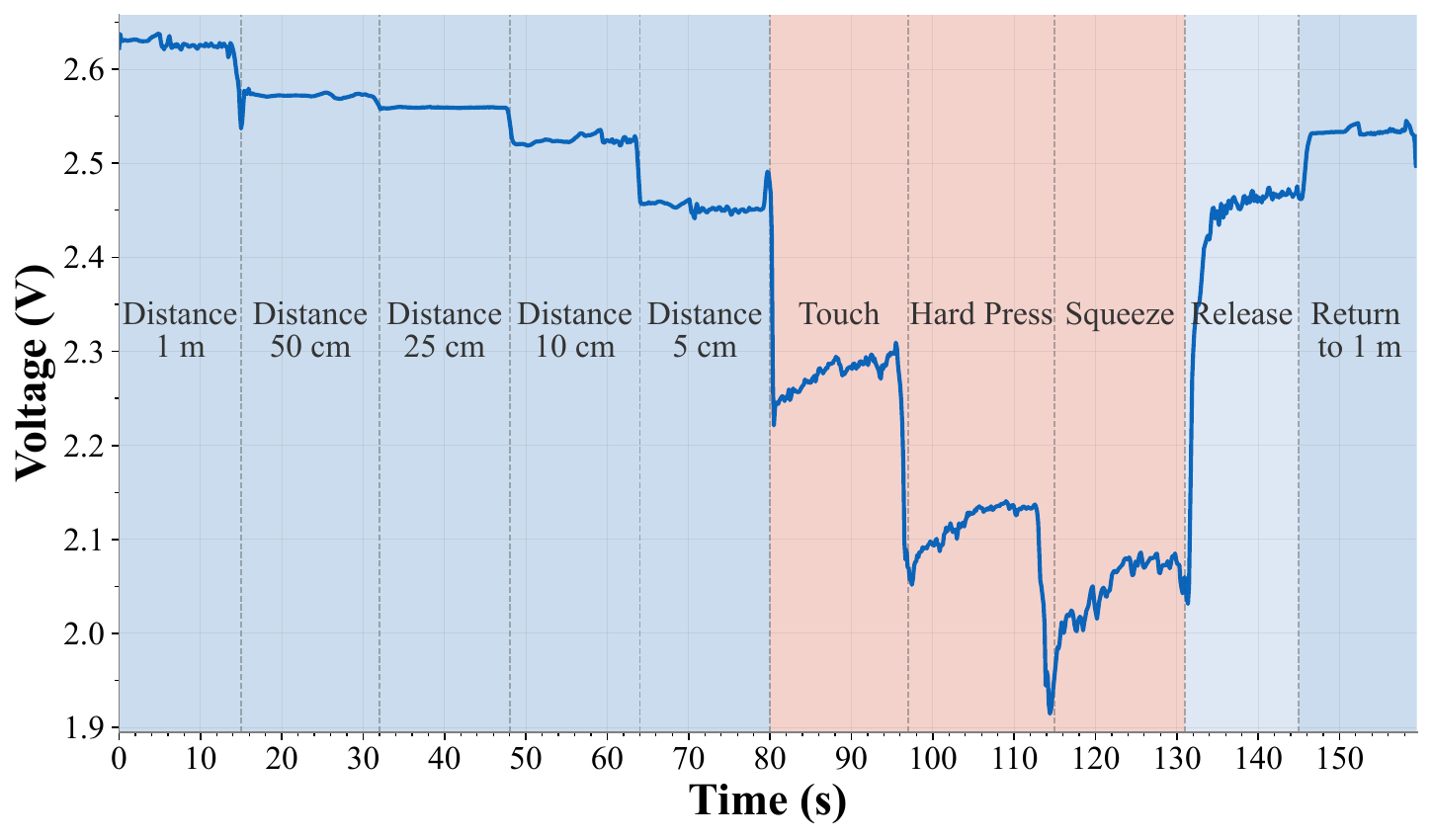}
    \caption{Voltage swings perceived by a SEth node due to nearby movement or manipulation of the silicone substrate. This example was recorded for a scenario in which the test subject moves perpendicularly towards the testbed, eventually touching it at a distance $L = 150$\,cm from the transmitter.}
    \label{fig:staircase}
\end{figure}

As shown in Figure~\ref{fig:staircase}, sudden changes in proximity to the testbed produce voltage swings up to hundreds of millivolts, which is an order of magnitude larger than signal variations caused by environmental noise. Movement up to one metre away can be reliably detected, and movements up to ten centimetres away can be tracked with centimetre-level accuracy. The size of movement-induced voltage swings dramatically increases as the distance between silicone substrate and test subject decreases, offering increased resolution. This effect continues when users squeeze rather than just touch the silicone skin, enabling fine-grained detection of relative changes in exerted pressure.

\begin{figure}
    \centering
    \includegraphics[width=0.90\linewidth]{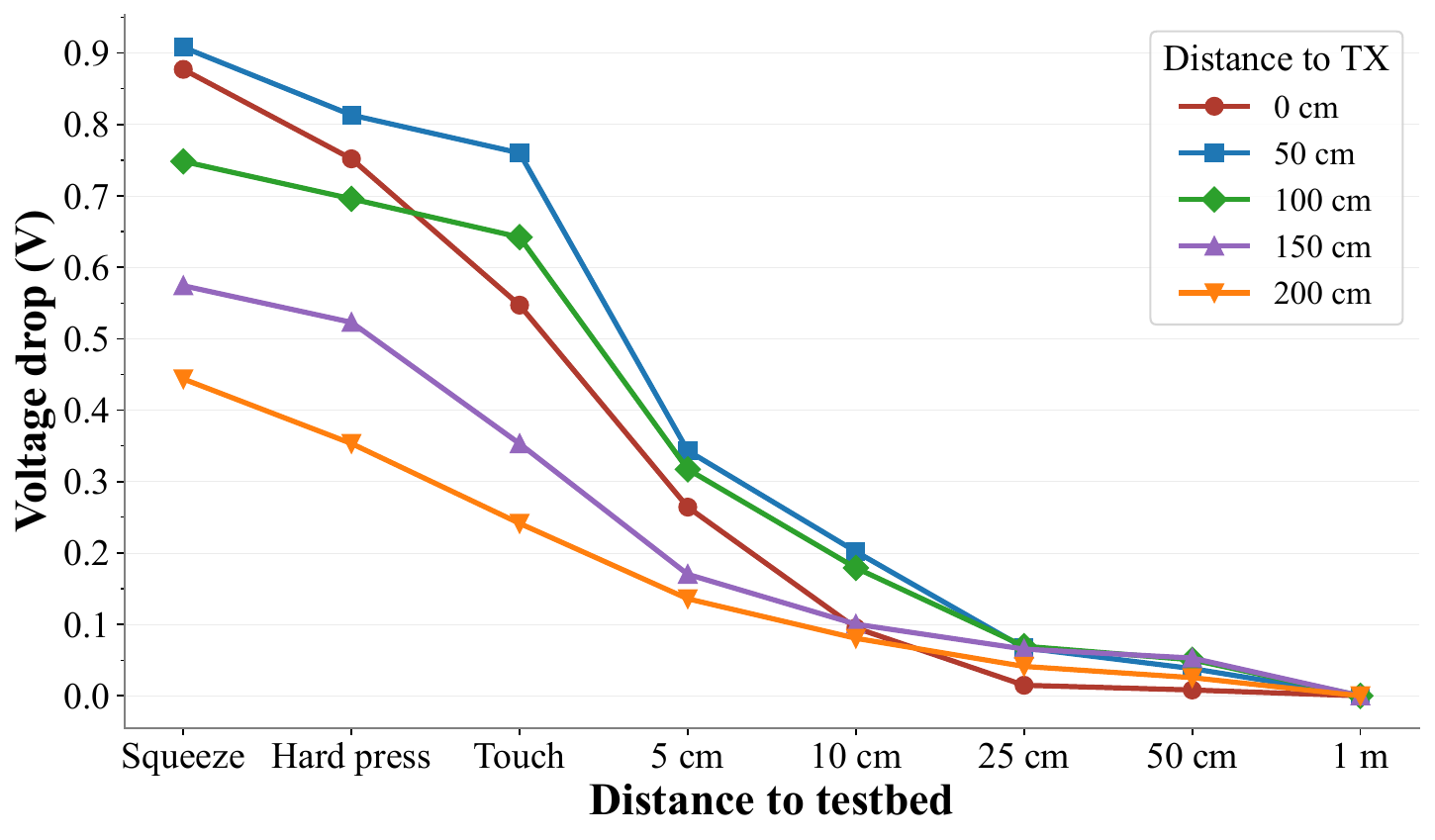}
    \caption{SEth detects centimetre-level movement up to circa 10\,cm away.}
    \label{fig:deltav}
\end{figure}

Figure~\ref{fig:deltav} shows the magnitude of movement-induced change in capacitor voltage for different positions $L$ along the axis between a SEth transmitter and receiver, relative to a setting in which the test subject is standing one metre away. We note that, when the test subject interacts with the SEth testbed from a distance $<10$ centimetres, the observed signal is significantly affected by $L$, and that the corresponding trend cannot be straightforwardly explained as a monotonic function of $L$. That is, close to the testbed, voltage measurements capture a combination movement parallel and from/towards the silicone surface. We hence believe that future work on inference over dense deployments of SEth nodes embedded within a silicone substrate could enable fine-grained localisation and interpretation of touch-based interactions (cf.~\cite{8272325}).

\subsection{Discussion and Limitations}
\label{sec:limitations}
To the best of our knowledge, SEth is the first example of a system which integrates communication, power transfer and sensing in a single wireless transceiver. The truly wireless and battery-free nature of SEth means that sensors can be deployed freely within the conductive silicone substrate, maximizing design flexibility and minimizing complexity. 

Considering network performance; SEth offers a compelling performance tradeoff with; 100\,kbps application-level throughput, high reliability, low latency and minimal jitter. Furthermore, SEth's non-destructive arbitration scheme for the first time enables real-time and priority-based channel sharing of a wireless link. 

Considering energy, the SEth transceiver is extremely low power. mW-scale transmitter power and sub-$\mu$W active listening eliminate the need for latency-inducing radio duty cycling, while efficient wireless power transfer can provide sufficient energy for 10s of messages per second at 1\,m range. It should also be noted that data rates can easily be increased by amplifying the transmit power of the node which is transferring energy.

Despite the advantages enumerated above, SEth is subject to the following limitations that will be addressed in future work:
\begin{enumerate}
    \item \emph{End-to-end network protocol:} In its current form the SEth transceiver essentially operates in one of three modes: communication, power transfer or sensing. A complete protocol has yet to be developed to orchestrate how nodes switch between these states.
    \item \emph{Evaluation at scale:} A key advantage of SEth is that its lack of cabling supports dense deployments of sensors, however our current evaluation is limited to a maximum of 10 nodes. Larger scale validation is naturally required to validate this advantage.
    \item \emph{Sensor characterization:} In this paper, we provide only a basic validation of SEth's touch sensing capability. A deeper characterization of this sensor is required to understand its performance and accuracy compared to conventional touch sensing solutions.
    \item \emph{Reliability in real-world scenarios} has yet to be tested. Given the novelty of SEth, it is important to explore its immunity (or lack thereof) to external interference sources. 
\end{enumerate}

\section{Related Work}
\label{sec:relatedwork}

\noindent \textbf{Positioning.} Wireless robotic materials embed densely deployed, miniaturised electronics within structural materials to create highly functional composites~\cite{10.1145/3131672.3131702}. Their lack of wired and mechanical connections increases sensor data quality and avoids the mechanical constraints of e.g.\ wireframes, but in turn necessitates innovations regarding (1) communication between embedded nodes and (2) power transfer, (3) sensing, and (4) actuation~\cite{10.1145/3131672.3131702, mcevoy2015materials}. This paper provides a joint solution for problems (1) through (3), establishing a fieldbus-like network built on capacitive coupling between devices within a somewhat-conductive silicone substrate. We note that not all wireless robotic materials jointly tackle the same set of challenges. Beamformed signals that locally heat materials can, for example, cause localised actuation in much larger materials, without providing a generic communication platform~\cite{10.1145/3570361.3592494}.

\textbf{Physical structure}. A key difficulty in the physical design of wireless robotic materials lies in the structural impact of the antennas necessary for communication~\cite{pyo2021recent}. Researchers have hence argued for terahertz networking as a possible enabler for intra-soft-robot communication, because such high frequencies enable extremely small antennas~\cite{lemic2021survey}. Likewise, backscatter-based approaches can, to some extent, avoid the structural impact of supporting radio-frequency carrier generation at every node~\cite{10.5555/3324320.3324403}. Communication through near-field coupling, examined in this paper, similarly avoids the need for radiative antennas~\cite{nie2019textile}, thereby minimizing size and mechanical complexity.

The conductive silicone rubber on which this paper relies has similar electrical properties as living tissue. Near-field, capacitively coupled communication has previously been shown to support ultra-low-power on- or in-body networks~\cite{gu20242m}. This paper shows that order-of-magnitude performance improvements can be achieved in controlled media rather than the human body when controlling for range (cf. the bus network in~\cite{liu2025enoch}). Still, several authors report much faster capacitively-coupled on-body networks in low-noise or short-range settings (e.g.\ subcutaneous implants )~\cite{8805102}. The electrical properties of structural materials have also been exploited for communication between rather than within robots: Yang et al.\ describe an approach wherein ``origami'' conductive materials function as antennas at Ultra High Frequencies~\cite{yang2019multifunctional}.

\textbf{Logical structure.} To the best of our knowledge, this paper is the first to examine near-field coupling as a means of \emph{digital} communication in wireless robotic materials. In the analog domain, capacitive coupling is one of many well-known techniques for sensing and power transfer~\cite{nie2019textile}, with alternatives such as acoustic sensing and transfer~\cite{10.1145/3718958.3750486, 9561898, rajendran2024single}, radio-frequency energy harvesting~\cite{zhang2024skin}, inductive coupling~\cite{10105616}, and  optical power transfer and sensing~\cite{garcia2019energy, budelmann2013sensorial}. This paper reconciles such analog effects with the digital communication requirements of relevant applications. For example, the interpretation of touch via dense pressure sensor arrays requires tens of kbps of throughput and communication latency on the order of 10\,ms for control loops involving tens of nodes~\cite{9439200, 8272325}. We note that these latency requirements preclude the use of conventional wireless technologies such as Bluetooth \cite{tosi2017performance}  (cf.~\cite{markvicka2020wireless}). Baca et al.\ have previously proposed an intra-robot communication solution for latency-sensitive actuation that, similar to this paper, realises a fieldbus network ~\cite{5692368} (i.e.\ CAN~\cite{farsi1999overview}), but their solution is limited to modular (i.e.\ segmented) rather than ``soft'' robotics.  Likewise, acoustic backscatter networks, which similarly to this paper combine energy transfer and digital communication, fall short of this performance requirement by an order of magnitude and depend on rigid communication media~\cite{10.1145/3718958.3750486}.

\textbf{Materials and inference.} The physical implementation of the system detailed in this paper may be improved considerably in the future: the design of conductive elastomers is a research subject on its own~\cite{chen2017stretchable}, and so are the design and optimisation of integrated circuits for capacitively coupled communication~\cite{gu20242m}, as well as the mass production of embedded electronics on flexible substrates~\cite{liu2022neuro}. Several authors also argue that the inference over increasingly densely deployed sensors in robotic skin requires communication networks to mimic biological neural networks~\cite{liu2022neuro}, i.e. that the associated data volume necessitates in-network computation~\cite{hughes2015texture}. Multi-hop wireless robotic materials~\cite{10.5555/3324320.3324403}  provide for computation at intermediate nodes as data propagates towards its destination. As arbitration-based contention resolution can be exploited to provide in-network computation too~\cite{andersson2008scalable}, our future work will examine to what extent that functionality can be exploited for efficient inference over sensor arrays.

\textbf{Building on prior work.} SEth builds on three pieces of our previous work. Static~\cite{static} is a proof-of-concept for VHF wireless power transfer using large antennas in free-space. Static achieved sufficient wireless power transfer to power heavily duty-cycled commodity radios at a range of up to 5 meters. CaIN~\cite{liu2024cain} uses a similar receiver design, which is reparameterised for fast communication instead of energy transfer and extended with hardware demodulation support to realize a free-space near-field coupled transceiver with an effective range of over 15m and speeds of up to 20\,kbps. EnOCH~\cite{liu2025enoch} adapts static for low-latency on-body communication at 12.5\,kbps using capacitive coupling. Elements of each system design can be seen in the RX and harvesting front ends of SEth, which combines both approaches with wireless sensing and moves to a promising new communication medium.

\section{Conclusion}
\label{sec:conclusion}
This paper introduced SEth, a nervous system for soft robots, that is composed of networks of `neurons' suspended in a conductive silicone medium. Each battery-free neuron combines sensing, computation, communication and energy harvesting.

While SEth is simple and low cost, evaluation shows that it provides a compelling performance envelope with; 100\,kbps application-level throughput, millisecond-scale latency, microsecond scale jitter and unique real-time networking features. The wireless power transfer feature of SEth enables reliable and maintenance-free operation without the need for power cables or batteries, enabling data rates of up to 27 messages per second at 0.5\,m distance.

In addition to communication and power transfer, the SEth transceiver also enables wireless touch, pressure, and proximity sensing at a range of up to 1\,m. To the best of our knowledge, this is the \emph{first example of simultaneous wireless power transfer, communication, and sensing using a single transceiver}.

In the event of publication, all hardware and software presented in this paper will be made available under an open-source license to promote replication of these results and adoption of the approach.

\section{Future Work}
 \label{sec:futurework}
Our immediate future research will focus on addressing the limitations discussed in Section~\ref{sec:limitations}. 

In terms of an~\emph{end-to-end network protocol}, we will draw inspiration from wired embedded bus models such as 1-wire, which provide framing and addressing along with logic to switch between power and data transmission on a single transmission line. The 1-wire protocol could easily be transparently ported on top of SEth to provide an application layer, though extensions would be necessary to support a touch-sensing phase of interaction.

Additional characterization of touch sensing performance in large networks of SEth nodes is of great interest, particularly the accuracy of touch localization and its relationship to the range between SEth neurons. It is our belief that trilateration of touch sensing signals could be applied to provide accurate localization of touch using relatively sparse sensing arrays.

Large-scale testing and investigation of real-world reliability will be tackled by extending a commodity robot with a SEth enabled `skin' containing a minimum of 100 SEth neurons and deploying this robot in realistic settings for performance evaluation.

Beyond the near-term, we will investigate miniaturization of the SEth neuron. The current design is compact by the standards of typical radio prototypes, however, SEth offers unique opportunities to reach millimeter scale. While conventional radios require crystal oscillators, large antennas and bulky batteries, none of these components are necessary for SEth, which can be realized in pure silicon at millimeter scale.

\begin{acks}
This work is partially funded by the Research Fund, KU Leuven (C14/24/093; SwarmNet) and J. Oostvogels' Research Foundation - Flanders Fellowship (FWO; 1224325N).
\end{acks}

  \bibliographystyle{ACM-Reference-Format}
  \bibliography{sample-base}

@inproceedings{liu2024cain,
  title={{CaIN: Low Power and Low Latency VHF Mesh Networking}},
  author={Liu, Mengyao and Fang, Bingwu and Oostvogels, Jonathan and Michiels, Sam and Belogaev, Andrei and Liu, Xinlei and Famaey, Jeroen and Hughes, Danny},
  booktitle={EWSN'24: Proceedings of the 2024 International Conference on Embedded Wireless Systems and Networks},
  year={2024}
}

@ARTICLE{static,
AUTHOR={Thangarajan, Ashok Samraj  and Nguyen, Thien Duc  and Liu, Mengyao  and Michiels, Sam  and Yang, Fan  and Man, Ka Lok  and Ma, Jieming  and Joosen, Wouter  and Hughes, Danny }, 
TITLE={Static: Low Frequency Energy Harvesting and Power Transfer for the Internet of Things},
JOURNAL={Frontiers in Signal Processing},  
VOLUME={Volume 1 - 2021},
YEAR={2022},
URL={https://www.frontiersin.org/journals/signal-processing/articles/10.3389/frsip.2021.763299},
DOI={10.3389/frsip.2021.763299},
ISSN={2673-8198},
}

@book{horowitz1989art,
  title={The art of electronics},
  author={Horowitz, Paul and Hill, Winfield and Robinson, Ian},
  volume={2},
  year={1989},
  publisher={Cambridge university press Cambridge}
}

@standards{IEEE802.11,
  title={IEEE Standard for Information Technology--Telecommunications and Information Exchange between Systems Local and Metropolitan Area Networks--Specific Requirements Part 11: Wireless LAN Medium Access Control (MAC) and Physical Layer (PHY) Specifications},
  number={IEEE Std 802.11™-2024},
  publisher={{IEEE}},
  year={2024}
}

@standards{IEEE802.15.4,
  title={IEEE Standard for Information technology-- Local and metropolitan area networks-- Specific requirements-- Part 15.4: Wireless Medium Access Control (MAC) and Physical Layer (PHY) Specifications for Low Rate Wireless Personal Area Networks (WPANs)},
  number={IEEE Std 802.15.4-2020},
  publisher={{IEEE}},
  year={2020}
}

@standards{ISO11898,
  title = {{Road Vehicles - Controller Area Network (CAN) - Part 1: Data Link Layer and Physical Layer}},
  howpublished = {International Organization for Standardization},
  year = {1993}
}

@article{EtherCAT,
  title={Development of EtherCAT real-time control system for robot based on Simulink Real-Time},
  author={Author, A. and Author, B.},
  journal={Journal of Control Science and Engineering},
  volume={2021},
  year={2021},
  doi={10.3233/JCM-204325}
}

@String{Computing = "Computing" }

@INPROCEEDINGS{9561898,
  author={Yang, Min Jin and Park, Kyungseo and Kim, Jung},
  booktitle={2021 IEEE International Conference on Robotics and Automation (ICRA)}, 
  title={A Large Area Robotic Skin with Sparsely Embedded Microphones for Human-Robot Tactile Communication}, 
  year={2021},
  volume={},
  number={},
  pages={3248-3254},
  doi={10.1109/ICRA48506.2021.9561898}}

@article{budelmann2013sensorial,
  title={From sensorial to smart materials: intelligent optical sensor network for embedded applications},
  author={Budelmann, Christoph and Krieg-Br{\"u}ckner, Bernd},
  journal={Journal of intelligent material systems and structures},
  volume={24},
  number={18},
  pages={2183--2188},
  year={2013},
  publisher={Sage Publications Sage UK: London, England}
}

@article{tosi2017performance,
  title={Performance evaluation of bluetooth low energy: A systematic review},
  author={Tosi, Jacopo and Taffoni, Fabrizio and Santacatterina, Marco and Sannino, Roberto and Formica, Domenico},
  journal={Sensors},
  volume={17},
  number={12},
  pages={2898},
  year={2017},
  publisher={MDPI}
}

@ARTICLE{8272325,
  author={Hughes, Dana and Lammie, John and Correll, Nikolaus},
  journal={IEEE Robotics and Automation Letters}, 
  title={A Robotic Skin for Collision Avoidance and Affective Touch Recognition}, 
  year={2018},
  volume={3},
  number={3},
  pages={1386-1393},
  doi={10.1109/LRA.2018.2799743}}

@inproceedings{liu2025enoch,
  title={ENOCH: ENabling On-body network Contention Handling},
  author={Liu, Mengyao and Oostvogels, Jonathan and Fang, Bingwu and Michiels, Sam and Ma, Haoxiang and Yang, Yang and Hughes, Danny},
  booktitle={2025 IEEE International Conference on Pervasive Computing and Communications (PerCom)},
  pages={121--127},
  year={2025},
  organization={IEEE}
}

@article{cianchetti2018biomedical,
  title={Biomedical applications of soft robotics},
  author={Cianchetti, Matteo and Laschi, Cecilia and Menciassi, Arianna and Dario, Paolo},
  journal={Nature Reviews Materials},
  volume={3},
  number={6},
  pages={143--153},
  year={2018},
  publisher={Nature Publishing Group UK London}
}

@ARTICLE{9829727,
  author={Elfferich, Johannes F. and Dodou, Dimitra and Santina, Cosimo Della},
  journal={IEEE Access}, 
  title={Soft Robotic Grippers for Crop Handling or Harvesting: A Review}, 
  year={2022},
  volume={10},
  number={},
  pages={75428-75443},
  doi={10.1109/ACCESS.2022.3190863}}

@INPROCEEDINGS{5692368,
  author={Baca, Jose and Ferre, Manuel and Collar, Matias and Fernandez, Jose and Aracil, Rafael},
  booktitle={2010 IEEE Electronics, Robotics and Automotive Mechanics Conference}, 
  title={Synchronizing a Modular Robot Colony for Cooperative Tasks Based on Intra-Inter Robot Communications}, 
  year={2010},
  volume={},
  number={},
  pages={388-393},
  doi={10.1109/CERMA.2010.111}}

@ARTICLE{farsi1999overview,
  author={M. Farsi and K. Ratcliff and M. Barbosa},
  journal={Computing \& Control Engineering Journal}, 
  title={An overview of {Controller Area Network}}, 
  year={1999},
  volume={10},
  number={3},
  pages={113--120},
  publisher={IETF}
}

@ARTICLE{9439200,
  author={Gbouna, Zakka Vincent and Pang, Gaoyang and Yang, Geng and Hou, Zeyang and Lyu, Honghao and Yu, Zhangwei and Pang, Zhibo},
  journal={IEEE Journal of Biomedical and Health Informatics}, 
  title={User-Interactive Robot Skin With Large-Area Scalability for Safer and Natural Human-Robot Collaboration in Future Telehealthcare}, 
  year={2021},
  volume={25},
  number={12},
  pages={4276-4288},
  doi={10.1109/JBHI.2021.3082563}}

@article{mcevoy2015materials,
  title={Materials that couple sensing, actuation, computation, and communication},
  author={McEvoy, Michael Andrew and Correll, Nikolaus},
  journal={Science},
  volume={347},
  number={6228},
  pages={1261689},
  year={2015},
  publisher={American Association for the Advancement of Science}
}

@inproceedings{10.1145/3718958.3750486, author = {Wang, Weiguo and He, Yuan and Xie, Yadong and Xie, Chuyue and Kai, Yi and Hu, Chengchen}, title = {Acoustic Backscatter Network for Vehicle Body-in-White}, year = {2025}, isbn = {9798400715242}, publisher = {Association for Computing Machinery}, address = {New York, NY, USA}, url = {https://doi.org/10.1145/3718958.3750486}, doi = {10.1145/3718958.3750486}, booktitle = {Proceedings of the ACM SIGCOMM 2025 Conference}, pages = {1040–1055}, numpages = {16}, location = {S\~{a}o Francisco Convent, Coimbra, Portugal}, series = {SIGCOMM '25} }

@article{zhang2024skin,
  title={Skin-inspired, sensory robots for electronic implants},
  author={Zhang, Lin and Xing, Sicheng and Yin, Haifeng and Weisbecker, Hannah and Tran, Hiep Thanh and Guo, Ziheng and Han, Tianhong and Wang, Yihang and Liu, Yihan and Wu, Yizhang and others},
  journal={Nature Communications},
  volume={15},
  number={1},
  pages={4777},
  year={2024},
  publisher={Nature Publishing Group UK London}
}

@ARTICLE{10105616,
  author={Dingley, Gavin and Cox, Mackenzie and Soleimani, Manuchehr},
  journal={IEEE Transactions on Instrumentation and Measurement}, 
  title={EM-Skin: An Artificial Robotic Skin Using Magnetic Inductance Tomography}, 
  year={2023},
  volume={72},
  number={},
  pages={1-9},
  doi={10.1109/TIM.2023.3268481}}

@article{garcia2019energy,
  title={Energy autonomous electronic skin},
  author={Garc{\'\i}a N{\'u}{\~n}ez, Carlos and Manjakkal, Libu and Dahiya, Ravinder},
  journal={npj Flexible Electronics},
  volume={3},
  number={1},
  pages={1},
  year={2019},
  publisher={Nature Publishing Group UK London}
}

@article{rajendran2024single,
  title={Single and bi-layered 2-D acoustic soft tactile skin (AST2)},
  author={Rajendran, Vishnu and Parsons, Simon and others},
  journal={arXiv preprint arXiv:2401.14292},
  year={2024}
}

@ARTICLE{8057758,
  author={Raza, Mohsin and Aslam, Nauman and Le-Minh, Hoa and Hussain, Sajjad and Cao, Yue and Khan, Noor Muhammad},
  journal={IEEE Communications Surveys \& Tutorials}, 
  title={A Critical Analysis of Research Potential, Challenges, and Future Directives in Industrial Wireless Sensor Networks}, 
  year={2018},
  volume={20},
  number={1},
  pages={39-95},
  keywords={Wireless sensor networks;Standards;Protocols;Tutorials;Energy harvesting;Industries;Automation;Automation;IEEE802.15.4e;energy harvesting;IWSNs;ISA100.11a;MAC;TDMA;WirelessHART;WSNs;Zigbee;6LoWPAN;CSMA/CA;Castalia;Fieldbus;Ethernet;mesh;tree;star;bus;flat architecture;hierarchical;security;congestion;Wi-Fi;Bluetooth;UWB;wasp mote;re-mote;openmote;SunSPOT;OMNeT++;MF;MiXiM;OPNET;gradient;flat;proactive routing;data centric;PV;VLC;IoT;cognitive sensor networks;6lo WG;OpenWSN},
  _doi={10.1109/COMST.2017.2759725}}

@article{liu2022neuro,
  title={Neuro-inspired electronic skin for robots},
  author={Liu, Fengyuan and Deswal, Sweety and Christou, Adamos and Sandamirskaya, Yulia and Kaboli, Mohsen and Dahiya, Ravinder},
  journal={Science robotics},
  volume={7},
  number={67},
  pages={eabl7344},
  year={2022},
  publisher={American Association for the Advancement of Science}
}

@article{802154,
title = {{IEEE standard for low-rate wireless networks}},
journal = {IEEE Std 802.15.4-2020 (Revision of IEEE Std 802.15.4-2015)},
year = {2020},
numpages = {801},
_url = {https://doi.org/10.1109/IEEESTD.2020.9144691},
_doi = {10.1109/IEEESTD.2020.9144691},
}

@inproceedings{10.1145/1031495.1031508, author = {Polastre, Joseph and Hill, Jason and Culler, David}, title = {Versatile low power media access for wireless sensor networks}, year = {2004}, isbn = {1581138792}, publisher = {Association for Computing Machinery}, address = {New York, NY, USA}, url = {https://doi.org/10.1145/1031495.1031508}, doi = {10.1145/1031495.1031508},  booktitle = {Proceedings of the 2nd International Conference on Embedded Networked Sensor Systems}, pages = {95–107}, numpages = {13}, location = {Baltimore, MD, USA}, series = {SenSys '04} }

@inproceedings{markvicka2020wireless,
  title={Wireless electronic skin with integrated pressure and optical proximity sensing},
  author={Markvicka, Eric J and Rogers, Jonathan M and Majidi, Carmel},
  booktitle={2020 IEEE/RSJ International Conference on Intelligent Robots and Systems (IROS)},
  pages={8882--8888},
  year={2020},
  organization={IEEE}
}

@article{chen2017stretchable,
  title={Stretchable conductive elastomer for wireless wearable communication applications},
  author={Chen, Zhibo and Xi, Jingtian and Huang, Wei and Yuen, Matthew MF},
  journal={Scientific reports},
  volume={7},
  number={1},
  pages={10958},
  year={2017},
  publisher={Nature Publishing Group UK London}
}

@article{andersson2008scalable,
  title={A scalable and efficient approach for obtaining measurements in {CAN}-based control systems},
  author={Andersson, Bj{\"o}rn and Pereira, Nuno and Elmenreich, Wilfried and Tovar, Eduardo and Pacheco, Filipe and Cruz, Nuno},
  journal={IEEE Transactions on Industrial Informatics},
  volume={4},
  number={2},
  pages={80--91},
  year={2008},
  publisher={IEEE}
}

@article{hughes2015texture,
  title={Texture recognition and localization in amorphous robotic skin},
  author={Hughes, Dana and Correll, Nikolaus},
  journal={Bioinspiration \& biomimetics},
  volume={10},
  number={5},
  pages={055002},
  year={2015},
  publisher={IOP Publishing}
}

@inproceedings{10.1145/3570361.3592494, author = {Wang, Jingxian and Song, Yiwen and Zadan, Mason and Shen, Yuyi and Chen, Vanessa and Majidi, Carmel and Kumar, Swarun}, title = {Wireless Actuation for Soft Electronics-free Robots}, year = {2023}, isbn = {9781450399906}, publisher = {Association for Computing Machinery}, address = {New York, NY, USA}, url = {https://doi.org/10.1145/3570361.3592494}, doi = {10.1145/3570361.3592494}, booktitle = {Proceedings of the 29th Annual International Conference on Mobile Computing and Networking}, articleno = {2}, numpages = {16}, location = {Madrid, Spain}, series = {ACM MobiCom '23} }

@inproceedings{10.1145/3131672.3131702, author = {Correll, Nikolaus and Dutta, Prabal and Han, Richard and Pister, Kristofer}, title = {Wireless Robotic Materials}, year = {2017}, isbn = {9781450354592}, publisher = {Association for Computing Machinery}, address = {New York, NY, USA}, url = {https://doi.org/10.1145/3131672.3131702}, doi = {10.1145/3131672.3131702}, booktitle = {Proceedings of the 15th ACM Conference on Embedded Network Sensor Systems}, articleno = {24}, numpages = {6}, location = {Delft, Netherlands}, series = {SenSys '17} }

@inproceedings{10.5555/3324320.3324403, author = {Piumwardane, Dilushi and P\'{e}rez-Penichet, Carlos and Rohner, Christian and Voigt, Thiemo}, title = {Backscatter Communication for Wireless Robotic Materials}, year = {2019}, isbn = {9780994988638}, publisher = {International Conference on Embedded Wireless Systems and Networks (EWSN)}, booktitle = {Proceedings of the 2019 International Conference on Embedded Wireless Systems and Networks}, pages = {336–340}, numpages = {5}, location = {Beijing, China}, series = {EWSN '19} }

@article{lemic2021survey,
  title={Survey on terahertz nanocommunication and networking: a top-down perspective},
  author={Lemic, Filip and Abadal, Sergi and Tavernier, Wouter and Stroobant, Pieter and Colle, Didier and Alarc{\'o}n, Eduard and Marquez-Barja, Johann and Famaey, Jeroen},
  journal={IEEE Journal on Selected Areas in Communications},
  volume={39},
  number={6},
  pages={1506--1543},
  year={2021},
  publisher={IEEE}
}

@article{yang2019multifunctional,
  title={Multifunctional metallic backbones for origami robotics with strain sensing and wireless communication capabilities},
  author={Yang, Haitao and Yeow, Bok Seng and Li, Zhipeng and Li, Kerui and Chang, Ting-Hsiang and Jing, Lin and Li, Yang and Ho, John S and Ren, Hongliang and Chen, Po-Yen},
  journal={Science Robotics},
  volume={4},
  number={33},
  pages={eaax7020},
  year={2019},
  publisher={American Association for the Advancement of Science}
}

@ARTICLE{8805102,
  author={Maity, Shovan and Chatterjee, Baibhab and Chang, Gregory and Sen, Shreyas},
  journal={IEEE Journal of Solid-State Circuits}, 
  title={BodyWire: A 6.3-pJ/b 30-Mb/s -30-dB SIR-Tolerant Broadband Interference-Robust Human Body Communication Transceiver Using Time Domain Interference Rejection}, 
  year={2019},
  volume={54},
  number={10},
  pages={2892-2906},
  doi={10.1109/JSSC.2019.2932852}}

@article{gu20242m,
  title={A 2m-Range 711$\mu$W Body Channel Communication Transceiver Featuring Dynamically-Sampling Bias-Free Interface Front End},
  author={Gu, Guanjie and Yang, Changgui and Zhao, Jian and Du, Sijun and Luo, Yuxuan and Zhao, Bo},
  journal={IEEE transactions on biomedical circuits and systems},
  year={2024},
  publisher={IEEE}
}

@article{nie2019textile,
  title={Textile-based wireless pressure sensor array for human-interactive sensing},
  author={Nie, Baoqing and Huang, Rong and Yao, Ting and Zhang, Yiqiu and Miao, Yihui and Liu, Changrong and Liu, Jian and Chen, Xinjian},
  journal={Advanced functional materials},
  volume={29},
  number={22},
  pages={1808786},
  year={2019},
  publisher={Wiley Online Library}
}

@article{pyo2021recent,
  title={Recent progress in flexible tactile sensors for human-interactive systems: from sensors to advanced applications},
  author={Pyo, Soonjae and Lee, Jaeyong and Bae, Kyubin and Sim, Sangjun and Kim, Jongbaeg},
  journal={Advanced Materials},
  volume={33},
  number={47},
  pages={2005902},
  year={2021},
  publisher={Wiley Online Library}
}

@article{sun_soft_2022,
	title = {A soft thumb-sized vision-based sensor with accurate all-round force perception},
	volume = {4},
	doi = {10.1038/s42256-021-00439-3},
	number = {2},
	journal = {Nature Machine Intelligence},
	author = {Sun, Huanbo and Kuchenbecker, Katherine J. and Martius, Georg},
	year = {2022},
	pages = {135--145}}

@inproceedings{deanleon_wholebody_2019,
	address = {Montreal, QC, Canada},
	title = {Whole-{Body} {Active} {Compliance} {Control} for {Humanoid} {Robots} with {Robot} {Skin}},
	doi = {10.1109/ICRA.2019.8793258},
	booktitle = {2019 {International} {Conference} on {Robotics} and {Automation} ({ICRA})},
	publisher = {IEEE},
	author = {Dean-Leon, Emmanuel and Guadarrama-Olvera, J. Rogelio and Bergner, Florian and Cheng, Gordon},
	year = {2019},
	pages = {5404--5410}}

@inproceedings{ruppert_foottile_2020,
	title = {{FootTile}: a {Rugged} {Foot} {Sensor} for {Force} and {Center} of {Pressure} {Sensing} in {Soft} {Terrain}},
	doi = {10.1109/ICRA40945.2020.9197466},
	booktitle = {2020 {IEEE} {International} {Conference} on {Robotics} and {Automation} ({ICRA})},
	author = {Ruppert, Felix and Badri-Spr\"owitz, Alexander},
	year = {2020}}

@article{lee_nanomesh_2020,
	title = {Nanomesh pressure sensor for monitoring finger manipulation without sensory interference},
	volume = {370},
	doi = {10.1126/science.abc9735},
	number = {6519},
	journal = {Science},
	author = {Lee, Sunghoon and Franklin, Sae and Hassani, Faezeh Arab and Yokota, Tomoyuki and Nayeem, Md Osman Goni and Wang, Yan and Leib, Raz and Cheng, Gordon and Franklin, David W. and Someya, Takao},
	year = {2020},
	pages = {966--970}}

@inproceedings{kuchenbecker_characterizing_2003,
	address = {Washington, DC, USA},
	title = {Characterizing the {Human} {Wrist} for {Improved} {Haptic} {Interaction}},
	doi = {10.1115/IMECE2003-42017},
	booktitle = {Dynamic {Systems} and {Control}, {Volumes} 1 and 2},
	publisher = {ASMEDC},
	author = {Kuchenbecker, Katherine J. and Park, June Gyu and Niemeyer, Guenter},
	year = {2003},
	pages = {591--598}}

@article{mittendorfer_humanoid_2011,
	title = {Humanoid {Multimodal} {Tactile}-{Sensing} {Modules}},
	volume = {27},
	doi = {10.1109/TRO.2011.2106330},
	number = {3},
	journal = {IEEE Transactions on Robotics},
	author = {Mittendorfer, P and Cheng, G},
	year = {2011},
	pages = {401--410}}

@article{rus_design_2015,
	title = {Design, fabrication and control of soft robots},
	volume = {521},
	doi = {10.1038/nature14543},
	number = {7553},
	journal = {Nature},
	author = {Rus, Daniela and Tolley, Michael T.},
	year = {2015},
	pages = {467--475}}

@article{choi_vibrotactile_2013,
	title = {Vibrotactile {Display}: {Perception}, {Technology}, and {Applications}},
	volume = {101},
	doi = {10.1109/JPROC.2012.2221071},
	number = {9},
    journal = {Proceedings of the IEEE},
	author = {Choi, Seungmoon and Kuchenbecker, Katherine J.},
	year = {2013},
	pages = {2093--2104}}

@inproceedings{dietrich_development_2004,
	address = {Halle, Germany},
    year=2004,
	title = {Development of a peristaltically actuated device for the minimal invasive surgery with a haptic sensor array},
	isbn = {3-8322-2655-9},
	booktitle = {proceedings of the 2nd symposium: {Micro}-and {Nanostructures} of {Biological} {Systems}},
	publisher = {Shaker-Verlag},
	author = {Dietrich, Johannes and Meier, Petra and Oberthür, Siegfried and Preuß, Roman and Voges, Danja and Zimmermann, Klaus}}

@article{rogers_materials_2010,
	title = {Materials and {Mechanics} for {Stretchable} {Electronics}},
	volume = {327},
	doi = {10.1126/science.1182383},
	number = {5973},
	journal = {Science},
	author = {Rogers, John A. and Someya, Takao and Huang, Yonggang},
	year = {2010},
	pages = {1603--1607}}

@inproceedings{loher_stretchable_2006,
	address = {Singapore},
	title = {Stretchable electronic systems},
	doi = {10.1109/EPTC.2006.342728},
	booktitle = {2006 8th {Electronics} {Packaging} {Technology} {Conference}},
	author = {Loher, Thomas and Manessis, Dion and Heinrich, Ralf and Schmied, Benno and Vanfleteren, Jan and Debaets, Johan and Ostmann, Andreas and Reichl, Herbert},
	year = {2006},
	pages = {271--276}}

@article{grimminger_open_2020,
	title = {An {Open} {Torque}-{Controlled} {Modular} {Robot} {Architecture} for {Legged} {Locomotion} {Research}},
	volume = {5},
	doi = {10.1109/LRA.2020.2976639},
	language = {en},
	number = {2},
	journal = {IEEE Robotics and Automation Letters},
	author = {Grimminger, Felix and Meduri, Avadesh and Khadiv, Majid and Viereck, Julian and Wuthrich, Manuel and Naveau, Maximilien and Berenz, Vincent and Heim, Steve and Widmaier, Felix and Flayols, Thomas and Fiene, Jonathan and Badri-Spr\"owitz, Alexander and Righetti, Ludovic},
	month = apr,
	year = {2020},
	pages = {3650--3657}}

@article{sprowitz_roombots_2014,
	title = {Roombots: {A} hardware perspective on {3D} self-reconfiguration and locomotion with a homogeneous modular robot},
	volume = {62},
	doi = {10.1016/j.robot.2013.08.011},
	number = {7},
	journal = {Robotics and Autonomous Systems},
	author = {Spr\"owitz, Alexander and Moeckel, Rico and Vespignani, Massimo and Bonardi, Stephane and Ijspeert, Auke Jan},
	month = jul,
	year = {2014},
	pages = {1016--1033}}

@article{lu_flexible_2014,
	title = {Flexible and {Stretchable} {Electronics} {Paving} the {Way} for {Soft} {Robotics}},
	volume = {1},
	doi = {10.1089/soro.2013.0005},
	number = {1},
	journal = {Soft Robotics},
	author = {Lu, Nanshu and Kim, Dae-Hyeong},
	month = mar,
	year = {2014},
	pages = {53--62}}

@article{knight_energy_2008,
	title = {Energy {Options} for {Wireless} {Sensor} {Nodes}},
	volume = {8},
	doi = {10.3390/s8128037},
	language = {en},
	number = {12},
	journal = {Sensors},
	author = {Knight, Chris and Davidson, Joshua and Behrens, Sam},
	month = dec,
	year = {2008},
	pages = {8037--8066}}

@article{wang_biomorphic_2020,
	title = {Biomorphic structural batteries for robotics},
	volume = {5},
	issn = {2470-9476},
	doi = {10.1126/scirobotics.aba1912},
	number = {45},
	journal = {Science Robotics},
	author = {Wang, Mingqiang and Vecchio, Drew and Wang, Chunyan and Emre, Ahmet and Xiao, Xiongye and Jiang, Zaixing and Bogdan, Paul and Huang, Yudong and Kotov, Nicholas A.},
	year = {2020},
	pages = {eaba1912}}

@article{hepp_novel_2022,
	title = {A {Novel} {Spider}-{Inspired} {Rotary}-{Rolling} {Diaphragm} {Actuator} with {Linear} {Torque} {Characteristic} and {High} {Mechanical} {Efficiency}},
	volume = {9},
	doi = {10.1089/soro.2020.0108},
	number = {2},
	journal = {Soft Robotics},
	author = {Hepp, Jonas and Badri-Spr\"owitz, Alexander},
	year = {2022},
	pages = {364--375}}

@inproceedings{mockel_easy_2007,
	title = {An easy to use bluetooth scatternet protocol for fast data exchange in wireless sensor networks and autonomous robots},
	doi = {10.1109/IROS.2007.4399458},
	booktitle = {2007 {IEEE}/{RSJ} {International} {Conference} on {Intelligent} {Robots} and {Systems}},
	author = {Moeckel, Rico and Spr\"owitz, Alexander and Maye, Jerome and Ijspeert, Auke Jan},
	year = {2007},
	pages = {2801--2806}}

\end{document}